\documentclass[english,aps,prx,twocolumn,superscriptaddress]{revtex4-1}
\usepackage[english]{babel}
\usepackage{graphicx}
\usepackage{lipsum}
\usepackage[utf8]{inputenc}
\usepackage{amsmath}
\newcommand{\be}{\begin{equation}}
\newcommand{\ee}{\end{equation}}
\newcommand{\bea}{\begin{eqnarray}}
\newcommand{\eea}{\end{eqnarray}}
 
\usepackage{caption}
\usepackage{subfig}

\usepackage{color}
\usepackage[usenames,dvipsnames,svgnames,table]{xcolor}
\usepackage{hyperref}
\hypersetup{
     colorlinks   = true,
     citecolor    = red
}

\usepackage{xcolor}
\definecolor{ogreen} {RGB}{71,191,145}

\begin{document}
\title{Topological plasma oscillations in the solar tachocline}

\author{Ruben Lier}
\email{r.lier@uva.nl}
\affiliation{Institute for Theoretical Physics, University of Amsterdam, 1090 GL Amsterdam, The Netherlands}
\affiliation{Dutch Institute for Emergent Phenomena (DIEP), University of Amsterdam, 1090 GL Amsterdam, The Netherlands}

\author{Richard Green}
\email{rich.s.green@gmail.com}
\affiliation{Institute for Theoretical Physics, University of Amsterdam, 1090 GL Amsterdam, The Netherlands}

\author{Jan de Boer}
\email{J.deBoer@uva.nl}
\affiliation{Institute for Theoretical Physics, University of Amsterdam, 1090 GL Amsterdam, The Netherlands}
\affiliation{Dutch Institute for Emergent Phenomena (DIEP), University of Amsterdam, 1090 GL Amsterdam, The Netherlands}

\author{Jay Armas}
\email{j.armas@uva.nl}
\affiliation{Institute for Theoretical Physics, University of Amsterdam, 1090 GL Amsterdam, The Netherlands}
\affiliation{Dutch Institute for Emergent Phenomena (DIEP), University of Amsterdam, 1090 GL Amsterdam, The Netherlands}

\begin{abstract}
We study the properties of plasma oscillations in the solar tachocline using shallow-water magnetohydrodynamic equations. These oscillations are expected to correlate with solar activity. We find new qualitative features in the equatorial spectrum of magnetohydrodynamic oscillations associated with magneto-Rossby and magneto-Yanai waves. By studying this spectrum in terms of band theory, we find that magneto-Kelvin and magneto-Yanai waves are topologically protected. This highlights the important role of these two classes of waves, as robust features of the plasma oscillation spectrum, in the interpretation of helioseismological observations. 

\end{abstract}
\maketitle

\tableofcontents

\section{Introduction}
\begin{figure}[h!]
    \centering
    \includegraphics[width=8cm]{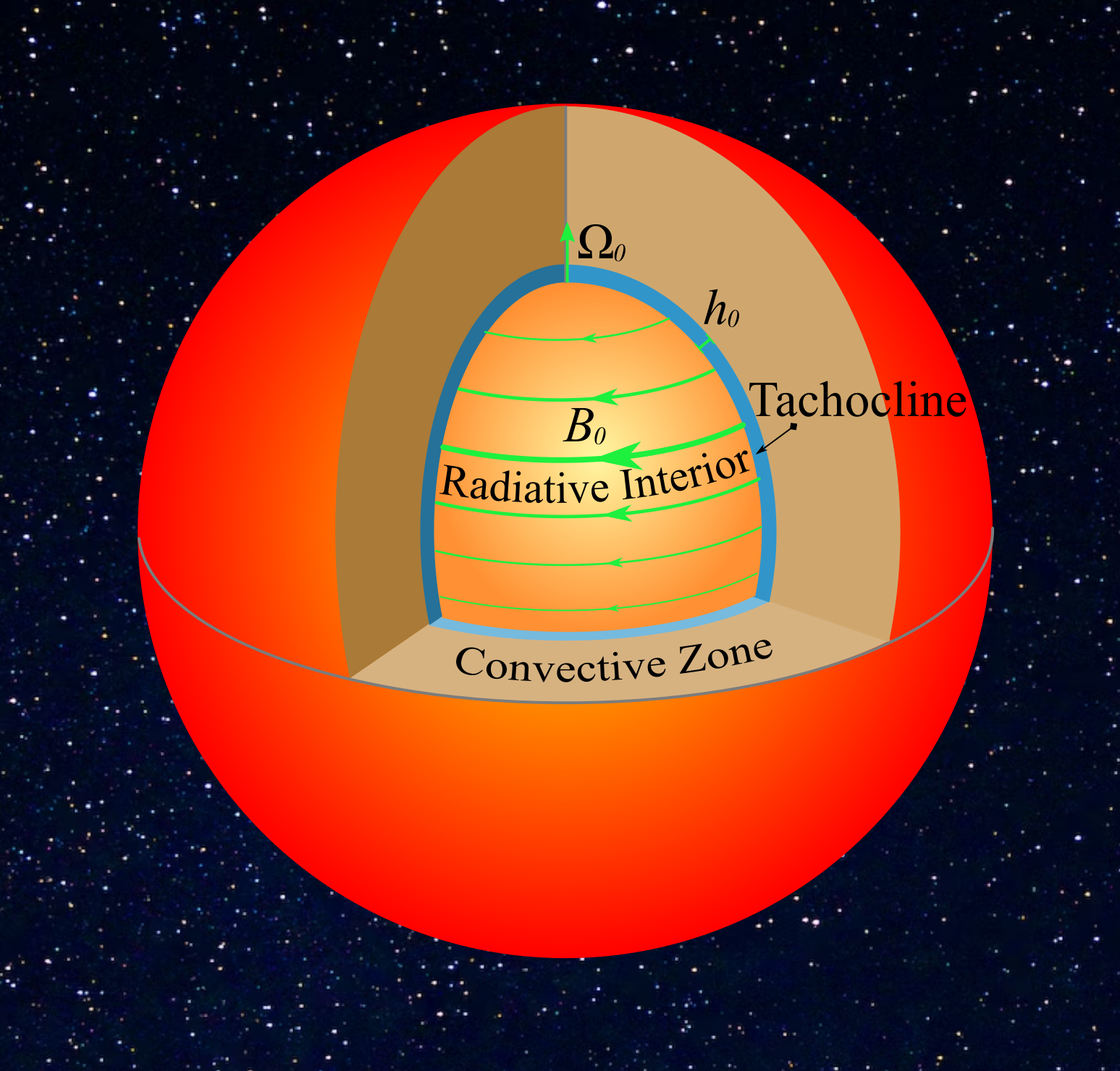}
\caption{Schematic picture of the Sun depicting the tachocline lying in between the radiative interior and the convective zone. Also displayed are three background variables which are used to describe the tachocline as a shallow water problem. These are the angular velocity $\Omega_0$ of the radiative interior, the height of the tachocline layer $h_0$ and the magnitude $B_0$ of the non-uniform toroidal magnetic field.}    \label{tachoclinepicture}
\end{figure}

In recent decades, helioseismology has revealed the internal rotation profile of the Sun. The inner radiative zone is rigidly rotating, while the outer convective zone has a non-trivial differential profile \cite{1989ApJ...343..526B, doi:10.1146/annurev.astro.41.011802.094848}.
Between the two zones is a thin layer - the solar tachocline - which marks the transition from rigid to differential rotation \cite{1992A&A...265..106S, Gilman_1997}. This layer is composed of a radiative part and an overshoot part deep in the convective zone at a distance of about 70\% of the solar radius and with a thickness of less than 5\% of the solar radius (see Fig.~\ref{tachoclinepicture}) \cite{Charbonneau_1999}. Current observations suggest that the tachocline does not change significantly in thickness or position with time \cite{Basu_2019}, while hosting strong toroidal magnetic fields \cite{hughes_rosner_weiss_2007}. The dynamics of the solar tachocline is a subject of intense study (see \cite{strugarek2023dynamics} for a review).

The influence of the tachocline on solar activity is not yet fully understood, but it has been argued that it may play a significant role in a variety of phenomena, including solar cycles \cite{strugarek2023dynamics}. Indeed, it has been suggested that the long wavelength oscillations of the magnetised plasma confined in the tachocline may correlate with observed sunspot activity over time \cite{Zaqarashvili_2010, Zaqarashvili_2015}.  An interesting example is the propagation of Rossby waves, which are expected to play an important role in space weather prediction, similar to their crucial role in jet stream activity on Earth \cite{Dikpati_2020, 2021SSRv..217...15Z}.

The dynamics of the plasma within this thin layer is typically approximated by the shallow water magnetohydrodynamic equations \cite{Gilman_2000, Dikpati_2001, Schecter_2001}. Within this framework, the dynamics of the plasma along the transverse (radial) direction is assumed to be approximately constant, effectively reducing the problem to two spatial dimensions. This system of equations has been extensively studied by various analytical approaches and nonlinear numerical simulations, leading to the realisation that the tachocline can host a rich structure of magnetohydrodynamic (MHD) oscillations, including Kelvin, Yanai, Rossby and inertial-gravity waves \cite{Dikpati_2001, Schecter_2001, Dikpati_2003, 2007A&A...470..815Z, Zaqarashvili_2009, npg-20-893-2013, doi:10.1080/03091929.2017.1301937, Zaqarashvili_2018, Belucz_2023}. Of particular importance are Rossby waves which have recently been observed on the surface of the Sun (see e.g. \cite{NatureRossby, 2019A&A...622A.124A, Hanasoge_2019, Waidele_2023}).

Our work in this paper is concerned with a deeper understanding of the properties of this rich structure of MHD oscillations, and is motivated by recent developments in uncovering topological protection mechanisms in continuous classical hydrodynamic systems (see e.g. \cite{souslov2019topological,PhysRevX.7.031039,fossati2022odd,green2020topological, Tong:2022gpg, monteiro2023coastal}). The prime example of such phenomena is the topological protection of Kelvin and Yanai waves in the Earth's oceans and atmosphere \cite{Delplace1075, delplacereal, Venaille_2021}, modelled by non-magnetic shallow water equations, while recent work uncovered topological imprints in the context of stellar oscillations \cite{Leclerc_2022, leclerc2023exceptional}.
The existence of topological mechanisms is not only interesting, but also crucial for our understanding of the robust features of the oscillation spectrum that is ultimately expected to be inferred through asteroseismology observations. The inner structure of the Sun is a complex dynamical system, and the MHD shallow water equations modelling the tachocline are at best an approximation to the actual dynamics. Topological properties associated with a given system typically indicate that not too drastic deformations of the system, including deformations of the Hamiltonian (and hence of the dynamics at the tachocline) or deformations of the geometry of certain parts of the system (e.g. the equatorial plane of the solar tachocline), do not lead to changes of physical observables (e.g. the spectrum of MHD oscillations at the tachocline).

In light of the above, we begin in Section \ref{sec:MHDSW} with a review of the MHD shallow water equations while in Section \ref{sec:equatorialspectrum} we revisit the spectrum of equatorially trapped waves in the solar tachocline using the $\beta$-plane approximation in which the spatial dependence of the Coriollis force at the equator is taken into account. This analysis was first carried out in \cite{Zaqarashvili_2018} but, as we will show, we find important differences in the spectrum compared to previous literature. These differences have implications for the potential imprint of the tachocline on the physics behind certain solar cycles. In Section \ref{sec:topology} we consider the MHD shallow water spectrum from the point of view of band structure theory and its topological properties, in particular the Chern number associated with each band. To do this, 
we perform an analysis of the spectrum near the equator using the f-plane approximation in which the Corriolis force is assumed to abruptly change sign at the equator, as pioneered in \cite{delplacereal}. The main result is that magneto-Kelvin and magneto-Yanai waves are robust features of the MHD spectrum in the solar tachocline. In Section \ref{sec:discussion} we conclude with some observations and future directions. We also provide some appendices, including Appendix \ref{app:shallowaterequations} with a derivation of the MHD shallow water equations; Appendix \ref{app:zaqa} with a comparison with previous literature; Appendix \ref{app:constantB} with a discussion of the MHD equatorial spectrum with constant magnetic fields; and finally Appendix \ref{app:berrycurvature} with details on the calculation of the Berry curvature, the f-plane approximation at the equator, and further topological insights into the MHD shallow water system.

\section{Magnetic shallow water equations}
\label{sec:MHDSW}
The MHD shallow water equations can be derived by considering a three dimensional incompressible plasma bounded between a rigid (bottom) boundary and a (top) dynamical interface (see Appendix \ref{app:shallowaterequations} for a derivation). These equations in two spatial dimensions can be written in the form
 \begin{subequations} \label{eq:allequations}
      \begin{align}
      D_t  h  &  =-  h \nabla_{\mu} u^{\mu}  ~~ , \label{eq:height}  \\ \label{eq:viscosityyyy}
 D_t u^{\mu}    & =  - g  \nabla^{\mu}  \eta + \frac{1}{\mu_0 \rho} B^{\nu} \nabla_{\nu} B^{\mu}   \\ 
D_t B^{\mu}  &  =  B^{\nu} \nabla_{\nu} u^{\mu}   ~~ , \label{eq:faraday} \\ 
     \nabla_{\mu} (h B^{\mu})  &  =0  ~~ ,  \label{eq:gausslaw}
    \end{align}
 \end{subequations} 
 where we have introduced the operator $D_t = \partial_t  + u^{\mu} \nabla_{\mu}$, the covariant derivative $\nabla_\mu$ associated with the two dimensional spatial metric $g_{\mu\nu}$ with coordinates $X^\mu$, the height $h$ of the tachocline, $\eta$ the dynamical field that accounts for changes in height with respect to a latitudinally uniform reference point. Together with the topography $H$, which accounts for the oblateness of the rigid bottom of the tachocline, it forms the total height $h=\eta+H$. We also introduced the (constant) density $\rho\sim 200 kgm^{-3}$ of the plasma, the fluid velocity $u^\mu$, the magnetic field $B^\mu$, the magnetic permittivity $\mu_0$ and the effective acceleration of gravity $g$ in the tachocline. The Greek indices $\mu,\nu$ only run over two spatial directions. Eq.~\eqref{eq:gausslaw} descends from the Gauss law in three dimensions (see Appendix \ref{app:shallowaterequations}) and similarly to the three dimensional Gauss law it is not an independent equation. Indeed to see this we note that in general we find after some nontrivial algebra
\begin{align}
      D_t  \frac{ \nabla_{\mu} (h B^{\mu})}{h}  =0 ~~ , 
\end{align}
 upon using Eqs.~\eqref{eq:height} and \eqref{eq:faraday} as well as the symmetry properties of the Riemann tensor. Indeed this condition implies that either $D_t h\sim \nabla_{\mu} (h B^{\mu})$ and hence Eq.~\eqref{eq:gausslaw} is equivalent to \eqref{eq:height}, or $D_t \nabla_{\mu} (h B^{\mu})=0$ and hence Eq.~\eqref{eq:gausslaw} becomes a constraint equation that only needs to be satisfied at an initial Cauchy slice. For the configurations studied in this paper the latter condition holds. 

In the context of the solar tachocline, Eqs.~\eqref{eq:allequations} describe MHD flows on an approximately thin spherical shell with line element
\begin{align} \label{eq:metric1}
    ds^2=g_{\mu\nu}dX^\mu dX^\nu =  L^2 ( d \theta^2 + \sin^2 (\theta) d \phi^2  ) ~~ , 
\end{align}
and radius $L\sim 10^{9} m$, where $\phi$ is the longitudinal coordinate and $\theta$ the latitudinal coordinate. The strong magnetic field living on the tachocline is expected to be induced by the non-uniform profile of the toroidal rotation in the convective zone. In this picture, such motion from above would carry magnetic fields from the convective zone into stretched toroidal magnetic fields in the tachocline \cite{spiegelzahn, Gilman1997, Zaqarashvili_2018}. We thus focus on non-uniform toroidal magnetic fields which in equilibrium (see Fig.~\ref{tachoclinepicture}) take the form 
\begin{align} \label{eq:nonuniformmagneticfield}
    B^{\mu} =   \delta^{\mu}_{\phi} B_0 \sin(\theta) \cos(\theta) ~~ ,  \end{align}
where $\delta^{\mu}_{\nu}$ is the Kronecker delta and $B_0$ is typically of order $10^5$ G. In addition the plasma living on the tachocline is assumed to rotating with uniform angular velocity $\Omega_0\sim 2\times 10^{-6} s^{-1}$ inherited from the rigid motion of the radiative core. Thus in equilibrium we have $u^\mu=\delta^\mu_\phi \Omega_0$ and
 \begin{align} \label{eq:heigttt}
 \begin{split}
 & \eta - C = \\ 
&\frac{1 }{2 g  } \left( \Omega_0^2  \cos^2 ( \theta  )  
 - \frac{ B_0^2}{ \mu_0 \rho  }  \frac{3 \sin ^2(\theta ) \cos ^4(\theta ) + \cos ^6(\theta ) }{6} \right)   ~~,     
 \end{split}
\end{align}
where $C$ is an integration constant. We are interested in equilibrium solutions for which $h_0$ is uniform along the latitudinal direction. Hence, although $\eta_0$ is non-uniform along the latitudinal direction, the topography $H$ of the bottom of the tachocline can be chosen such that $h_0$ is uniform by appropriately using Eq.~\eqref{eq:heigttt}. This choice of $H$ can be understood as accounting for the oblateness of the Sun due to its rotation, in a similar way to accounting for the oblateness of the Earth to approximate the depth of the ocean as uniform along its latitude. It is clear that this equilibrium configuration for the tachocline solves all equations in \eqref{eq:allequations}. Our goal now is to obtain the spectrum of MHD waves under certain approximations. In the next section we focus on finding the spectra localised at the equator using the $\beta$-plane approximation while in the following section we study the spectra away from the equator as well as on the equator using the f-plane approximation in order to extract robust properties of magnetohydrodynamic waves.\\
    
\section{Equatorial spectrum with toroidal magnetic fields}
\label{sec:equatorialspectrum}
In order to obtain the spectrum of equatorial magnetohydrodynamic waves we fluctuate Eqs.~\eqref{eq:allequations} around the equilibrium configuration with constant magnetic field, angular velocity and height $h_0,\Omega_0, B_0$ respectively. We thus introduce arbitrary perturbations around the equilibrium state according to $h = h_0 + \delta h $, $u^{\mu} =  \delta^{\mu}_{\phi} \Omega_0 + \delta u^{\mu}$, and $B^{\mu} = \delta^{\mu}_{\phi} B_0 \sin(\theta) \cos(\theta) + \delta B^{\mu}$. By performing a boost to a rotating frame where $\phi \rightarrow \phi - \Omega_0 t$, Eqs.~\eqref{eq:allequations} become
\begin{subequations} \label{eq:combinaed}
       \begin{align}
         \left(   \partial_{t}    -  \mathcal{H}  \right)   \begin{bmatrix}
      \delta h&
          \delta u^{\phi }&
     \delta u^{\theta } &
   \delta   B^{\phi } &
    \delta  B^{\theta }
    \end{bmatrix}^{T} =0  ~~ , 
\end{align}
where the effective Hamiltonian $\mathcal{H}$ is given by
\begin{widetext}
\begin{align} \label{eq:bigmatrixxx}
  \mathcal{H} = \begin{bmatrix} 
        0 & -   h_0 \partial_\phi  & -  h_0 \partial_\theta  -  h_0 \cot(\theta )  &  0 & 0 \\  
       -  \frac{g}{ L^2 \sin^2  (\theta)}  \partial_{\phi } & 0 &   - 2   \Omega_0 \cot(\theta )  &  \frac{1}{\mu_0 \rho } B_0 l  \partial_{\phi }  &   \frac{2}{\mu_0 \rho }    B_0 \left(\frac{3}{2} \cos^2( \theta )  - \frac{1}{2} \sin^2(\theta)\right)      \\ 
        - \frac{g}{L^2}  \partial_{\theta } &  2    \Omega_0 l   & 0 &  -  \frac{2}{\mu_0 \rho }  B_0 l^2  & \frac{1}{\mu_0 \rho }  B_0 l  \partial_{\phi }  \\      0 & B_0 l  \partial_{\phi } &  B_0 (  \sin^2(\theta) - \cos^2(\theta)) & 0 & 0 \\       
                        0 & 0 & B_0 l  \partial_{\phi } & 0 & 0 \\       
        \end{bmatrix}  ~~  , 
 \end{align}
  \end{widetext}
  \end{subequations}  
 and where we defined $l =  \cos(\theta )\sin(\theta )$. We will now simplify Eq.~\eqref{eq:bigmatrixxx} by assuming that we are near enough to the equator so that $\sin^2(\theta) \approx 1$ and $\cos^2 (\theta) \approx  0 $. It is then possible to bring Eqs.~\eqref{eq:combinaed} to a more workable form via the rescaling of the coordinates and fluctuations according to 
 \begin{equation}  \label{rescalingssss}
    \begin{array}
    {@{}r@{\,=\,}lr@{\,=\,}l@{}}
  \begin{bmatrix} 
          \partial_{\phi  } \\ 
          \partial_{\theta   }
    \end{bmatrix}    &    \frac{2 }{\sqrt{\mathcal{G}} }    \begin{bmatrix}
          \partial_{x } \\ 
          \partial_{y  }
    \end{bmatrix}     &    \partial_t 
      &   2 \Omega_0 \partial_{\tau }  \\
   \begin{bmatrix}
          \delta B^{\phi } \\ 
             \delta B^{\theta }
    \end{bmatrix}    &        B_0   \begin{bmatrix}
          \delta B^{x } \\ 
             \delta B^{y  }
    \end{bmatrix}   &  \begin{bmatrix}
          \delta u^{\phi } \\ 
             \delta u^{\theta }
    \end{bmatrix}   &        \sqrt{\mathcal{G}} \Omega_0   \begin{bmatrix}
          \delta u^{x } \\ 
             \delta u^{y  }
    \end{bmatrix}    ~~ .
\end{array}
\end{equation}
We note that in Eq.~\eqref{rescalingssss} the (dimensionless) reduced gravity $\mathcal{G}=g h_0/(L^2\Omega_0^2)$ is taken to be approximately in the range $10^{-3}\le \mathcal{G}\le 10^{-1}$ in the overshoot part of the tachocline \cite{Zaqarashvili_2018}. Because the effective Hamiltonian \eqref{eq:bigmatrixxx} is independent of $\tau$ and $x$ as defined in \eqref{rescalingssss}, we can assume that solutions to \eqref{eq:combinaed} are of the form $\sim e^{i \omega \tau  - i k_x x }$ with frequency $\omega$ and momentum $k$. Introducing this ansatz in Eq.~\eqref{eq:combinaed} we find 
\begin{align}  \label{matrixrescaled11112}
 &   \begin{split}
 \begin{bmatrix}
         -  i \omega  &   i k_x   &  - \partial_y      +  f_1   &    0    & 0 \\  
          i k_x   & - i \omega  &    m   &     i k_x    m \gamma^2   &    f_2 
 \\ 
     -    \partial_y   &  -  m  &   - i \omega   &    0   &  i k_x m  \gamma^2     \\ 
             0 &  i k_x  m   & f_3    &   - i \omega  & 0  \\       
                        0 & 0 & i k_x m   & 0 & -  i \omega 
    \end{bmatrix}  \begin{bmatrix}
       \delta \hat{h}  \\ 
          \delta u^{x } \\ 
     \delta u^{y  }  \\ 
   \delta   B^{x  }  \\ 
    \delta  B^{y  }
    \end{bmatrix}  =0  ~~  ,     
  \end{split}
 \end{align} 
 where we have defined $  m   = - \cos(\theta )$. We have also introduced the ratio $ \gamma^2 = \frac{v_A^2 }{g h_0}$ where $v_A$ is the Alfv\'{e}n speed $v_A=B_0 L/\sqrt{\mu_0 \rho}$ and defined the rescaled $\delta \hat h$ fluctuation according to $\delta \hat{h}= \frac{\delta h}{h_0}$. We take $v_A=126m s^{-1}$ throughout this paper. In Eq.~\eqref{matrixrescaled11112} we have also introduced the functions
\begin{align} \label{eq:fterms}
\begin{split}
        f_1  & = \frac{\sqrt{\mathcal{G}}}{2} m  ~~ , ~~  f_2 =  -  \frac{1}{2} \gamma^2  \sqrt{\mathcal{G}}        ~~ ,  ~~  f_3  = \frac{1}{2} \sqrt{\mathcal{G}}  ~~ . 
        \end{split}
\end{align}
Because $L \gg \sqrt{g h_0}/\Omega_0 $ for the case of the Sun, terms involving $ \mathcal{G}$ are subleading and hence we are free to discard all terms in Eq.~\eqref{eq:fterms}. We note, however, that there is no practical obstruction in including them but their effect on the spectrum for small $ \mathcal{G}$ (as in the overshoot layer) is minimal and does not change the results qualitatively.
 
Since we are interested in the behavior near the equator, we use the $\beta$-plane approximation for both the Coriolis force as well as for the non-uniform magnetic field \cite{Zaqarashvili_2018}, that is, we expand $m$ as \begin{align}
        m \approx  \beta y   ~~ ,
 \end{align}
 where we have defined $\beta= \sqrt{\mathcal G}/2$ \footnote{Although we ignored the terms $f_1$, $f_2$, $f_3$ by assuming $\mathcal{G}$ to be small, we find that the solutions to Eq.~\eqref{fullsolutionsszzz} crucially depend on $\beta y $ because the beta-plane terms are the source of their decay at large $y$. Therefore, although the terms in Eq.~\eqref{eq:fterms} could safely be ignored, we must retain the beta-plane contributions.}. Under these assumptions it is possible to eliminate $\delta B^x$, $\delta B^y$ and $\delta u^x$ from Eq.~\eqref{matrixrescaled11112} and obtain a dimensionally reduced system of equations of the form
 \begin{subequations} \label{fullsolutionsszzz}
 \begin{align}
 \left(  \partial_y   -   \mathcal{Q} \right)    \begin{bmatrix}
          \delta u^{y  }  \\ 
            \delta \hat{h}  
    \end{bmatrix}  =0 ~~ ,     
 \end{align}
 where $\mathcal{Q}$ is given by the matrix
  \begin{align}  \label{matrixrescaled2398398}
  \begin{split}
&  \mathcal{Q} = \\  & \begin{bmatrix}
           \frac{k_x \omega  \beta y   }{\omega ^2-\gamma ^2 k_x^2 \beta^2 y^2 }   &   -  i \omega  + \frac{i \omega   k^2_x}{\omega ^2-\gamma ^2 k_x^2 \beta^2 y^2 } 
 \\ 
     -     i \omega   +    i \frac{k^2_x}{\omega}    \beta^2 y^2  \gamma^2  +  \frac{i \omega \beta^2 y^2  }{\omega ^2-\gamma ^2 k_x^2 \beta^2 y^2 }  &     \frac{ - \omega    k_x \beta y  }{\omega ^2-\gamma ^2 k_x^2 \beta^2 y^2 }   
    \end{bmatrix}   ~~  .   
  \end{split}
 \end{align}
    \end{subequations}
There are two types of solutions to Eq.~\eqref{fullsolutionsszzz}, namely the magneto-Kelvin solution and the quantum harmonic oscillator (QHO) solutions. We discuss these two possibilities in order.

\subsection{The magneto-Kelvin solution}
\begin{figure}
    \centering
    \includegraphics[width=8cm]{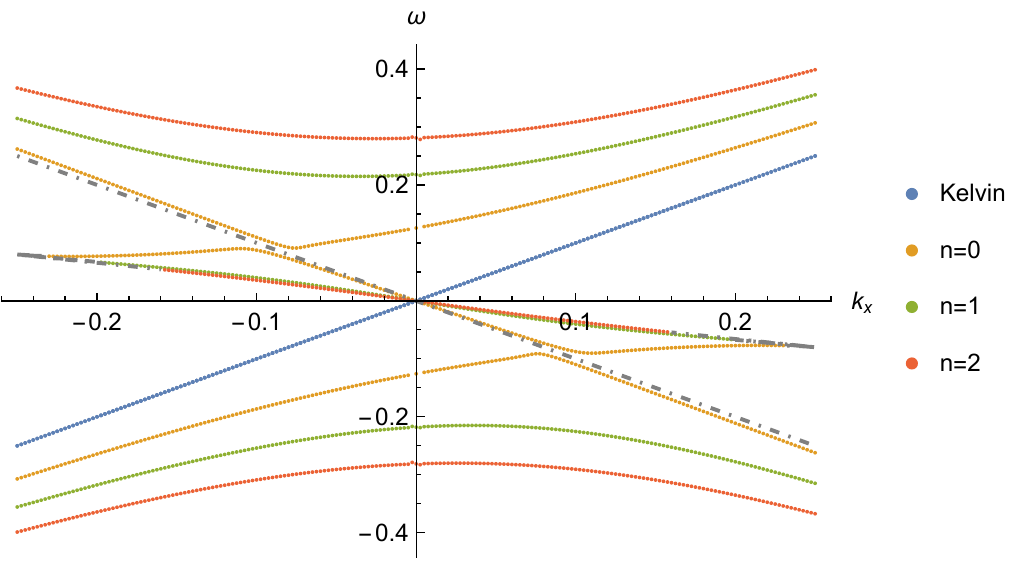}
\caption{Spectrum for equatorial magnetohydrodynamics with a non-uniform magnetic field including the magneto-Kelvin wave (blue curve) and the $n=0,1,2$ solutions of \eqref{eq:QHOsols}. The $n=0$ solutions (orange curves) include a slow magneto-Yanai wave passing through $\omega=0$ and a fast magneto-Yanai wave that asymptotes to the line $\omega=-k_x$ as $k_x\to-\infty$. The green and red curves correspond to the $n=1$ and $n=2$ solutions, respectively. For each $n\ge1$ we find a magneto-Rossby wave passing through $\omega=0$ and a magneto-inertial-gravity wave with $\omega\ne0$ at the origin $k_x=0$. The dashed lines correspond to (unphysical) non-normalizable modes. We used the values $\gamma= 1.56$ and $\mathcal{G} = 10^{-3}$. }    \label{fig:nonuniform}
\end{figure}

In order to find the magneto-Kelvin solution we expand Eq.~\eqref{fullsolutionsszzz} in powers of $y$ near the equator ($y=0$) and find
 \begin{align}  \label{matrixrescaled2398311198}
 &   \begin{split}
 \begin{bmatrix}
            \partial_y   - \beta y  \frac{ k_x     }{ \omega }   &   i \omega  -  \frac{i    k^2_x}{\omega } 
 \\           i \omega    &   \partial_y + \beta y \frac{k_x}{\omega}   
    \end{bmatrix}   \begin{bmatrix}
         \delta u^{y  }  \\ 
            \delta \hat{h} 
    \end{bmatrix}  = \mathcal{O} ( \beta^2  y^2 )  ~~  .   
  \end{split}
 \end{align}   
 Eq.~\eqref{matrixrescaled2398311198} admits a solution if $\omega = c k_x $, with $c=\pm 1$, and the fluctuations take the form
 \begin{align} \label{eq:outputtt}
      \begin{bmatrix}
           \delta u^{y  } \\ 
            \delta \hat{h} 
    \end{bmatrix}   =  \begin{bmatrix}
      0   \\ 
    \mathcal{C} \exp( - \frac{1}{2} c  \beta y^2  )  
    \end{bmatrix} ~~ ,
 \end{align}
 where $\mathcal{C}$ is an arbitrary constant. We require that this solution is normalizable when $y\to\infty$ which enforces that $c=1$. This ensures that the fluctation $\delta \hat h$ is exponentionally decaying away from the equator. The dispersion relation $\omega=k_x$ corresponds to the blue line in Fig.~\ref{fig:nonuniform}. 
The dispersion relation of the magneto-Kelvin wave is in fact the same as the Kelvin wave in ocean dynamics \cite{Matsuno1966}. In Fig.~\ref{fig:nonuniform} we have also plotted with a dashed line the (unphysical) non-normalizable mode $\omega=-k_x$.
\subsection{The quantum harmonic oscillator solutions}
Similarly to the case of vanishing magnetic field \cite{Matsuno1966}, this class of solutions includes an infinite tower of excitations. In order to obtain them, we map  Eq.~\eqref{fullsolutionsszzz} to a QHO oscillator problem in terms of some field $U$ determined by
\begin{align}  \label{eq:almostharmonic1}
     \partial_y^2   U    =  \left[  M  + N y^2      
   +  \mathcal{O} ( \beta^4  y^4 ) \right]     U   ~~  ,
      \end{align}     
      where $M$ and $N$ are coefficients that do not depend on $y$. To achieve this, we first solve Eq.~\eqref{fullsolutionsszzz} for $\delta \hat{h}$ and expand  around the equator ($y=0$) to obtain an equation of the form
\begin{align}  
\begin{split}
    \label{eq:almostharmonic}
     \partial_y^2  \delta u^y   &  =  \left[  A y^2   - B    +  \mathcal{O} ( \beta^4  y^4) \right]    \delta u^y    \\  & 
     +  2 y \left[ C + D y^2    
   +  \mathcal{O} ( \beta^4  y^4 )  \right]    \partial_y   \delta u^y   ~~  . 
   \end{split}
      \end{align}
Here the coefficients $A,B,C,D$ are given in terms of the frequency $\omega$, momentum $k$ and the physical parameters $\mathcal{G}$ and $\gamma$ according to
     \begin{equation} \label{eq:ABCDcoef}
         \begin{split}
                A        & = \frac{\sqrt{\mathcal{G}^3} \gamma ^2 k_x^3}{8 \omega ^3}+\frac{1}{4} \mathcal{G}  \left(\gamma ^2 k_x^2+1\right)-\frac{\sqrt{\mathcal{G}^3} \gamma ^2 k_x^3}{4 \omega  \left(k_x^2-\omega ^2\right)}~~ , 
     \\ 
          &  B   = \omega ^2 -k_x^2-\frac{\sqrt{\mathcal{G} } k_x}{2 \omega } ~~ , ~~C  = \frac{\mathcal{G}  \gamma ^2 k_x^4}{4 \omega ^2 (  k_x^2 - \omega ^2 ) }~~ ,   \\ 
           & D  = \frac{\mathcal{G} ^2 \gamma ^4 k_x^6}{16 \omega ^4 \left(k_x^2-\omega ^2\right)}-\frac{\mathcal{G} ^2 \gamma ^4 k_x^6}{8 \omega ^2 \left(k_x^2-\omega ^2\right)^2} ~~. 
           \end{split}
     \end{equation}
  Our goal is to recast Eq.~\eqref{eq:almostharmonic} with coefficients \eqref{eq:ABCDcoef} into the QHO form of Eq.~\eqref{eq:almostharmonic1}. To this end we define a fluctuation $\delta \tilde u^y$ according to
  \begin{align} \label{eq:transformation}
  \delta u^y = e^{ \frac{1}{2} C y^2  + \frac{1}{4} D y^4  + \mathcal{O} (y^6 )  } \delta  \tilde u^y  ~~ .       
  \end{align}
Using Eq.~\ref{eq:transformation} in Eq.~\eqref{eq:almostharmonic} we can bring it to the form \eqref{eq:almostharmonic1} such that
  \begin{align} \label{eq:finalequation}
     \partial_y^2  \delta  \tilde u^y  =  \left[( A + C^2 - 3 D ) y^2  - (B + C )  + \mathcal{O} ( \beta^4  y^4)  \right]   \delta  \tilde u^y  ~~  , 
      \end{align}
and hence we identify $U=\delta \tilde u^y$, $M=-(B+C)$ and $N=A+C^2-3D$. Given the QHO form of \eqref{eq:finalequation} it is straightforward to find solutions, which are given by the infinite tower of excitations
\begin{align} \label{eq:QHOsols}
    B + C  = (2 n +1 ) \sqrt{A + C^2 - 3 D} ~~ , ~~ n = 0,1,2... ~~ ,
\end{align}
for each value of $n$. We note the particular importance in taking into account the coefficient $D$ appearing at order $\mathcal{O}(y^3)$ in Eq.~\eqref{eq:almostharmonic} since it contributes at order $\mathcal{O}(y^2)$ in the QHO equation \eqref{eq:finalequation}.
 In terms of the actual fluctuations $\delta u^y$, the QHO solutions take the form
 \begin{align} \label{eq:uUrelation}
     \delta u^y \sim  e^{  \frac{1}{2} (   C  - \sqrt{A+ C^2 - 3 D } ) y^2  + \frac{1}{4} D y^4 +  \mathcal{O} ( \beta^6  y^6)   }  ~~ . 
 \end{align}
 Similarly to the analysis of the magneto-Kelvin wave we require QHO solutions for $\delta u^y$ to be bounded as $y\to\infty$ up to corrections of $\mathcal{O} (y^4) $. From Eq.~\eqref{eq:uUrelation} it thus follows that we must have
 \begin{align} \label{eq:stabilityconstr}
  \sqrt{A+ C^2 - 3 D }    -     C  \geq 0  ~~ , 
 \end{align}
 and we therefore discard solutions that violate Eq.~\eqref{eq:stabilityconstr}. 
 
 Finally, taking into account all these constraints and combining the Kelvin and QHO solutions, we find the spectrum for equatorial MHD waves in the solar tachocline portrayed in Fig.~\ref{fig:nonuniform}. This spectrum has the symmetry $(\omega,k_x)\to-(\omega,k_x)$.  The blue line appearing Fig.~\ref{fig:nonuniform} is the magneto-Kelvin wave solution with $\omega=k_x$ in Eq.~\eqref{eq:outputtt} while the dashed line with $\omega=-k_x$ is the (unphysical) non-normalizable solution of Eq.~\eqref{eq:outputtt}. The orange lines are the $n=0$ solutions given by Eq.~\eqref{eq:QHOsols} and referred to as magneto-Yanai waves. The slow magneto-Yanai wave approaches $\omega\to0$ for $k_x\to0$ and ends at a given value of $k_x<0$ beyond which it no longer satisfies the normalizability condition \eqref{eq:stabilityconstr} as indicated by the dashed lines. The fast magneto-Yanai wave approaches the line $w=-k_x$ as $k_x\to-\infty$ and approaches the magneto-Kelvin wave for $k_x\to\infty$. Both slow and fast magneto-Yanai waves are not chiral, meaning that they travel eastwards for certain values of $k_x$ and westwards for other values of $k_x$. In turn, for $n=1$ we find a magneto-Rossby wave, represented by the green curve in Fig.~\ref{fig:nonuniform} that approaches $\omega\to0$ for $k_x\to0$. This wave, as the slow magneto-Yanai wave, ends at a particular value of $k_x<0$ as indicated by the dashed lines that no longer satisfy \eqref{eq:stabilityconstr}. In addition, for $n=1$ there is a magneto-inertial-gravity wave, also portrayed as a green curve in Fig.~\ref{fig:nonuniform}, that has $\omega\ne0$ for $k_x=0$. The picture is similar for solutions of \eqref{eq:QHOsols} with $n\ge1$. The spectrum of equatorial MHD waves had been previous derived in \cite{Zaqarashvili_2018} and does not agree with the spectrum we derived here. We comment further on these differences in Appendix~\ref{app:zaqa}.
 
\begin{figure}
    \centering
    \includegraphics[width=8cm]{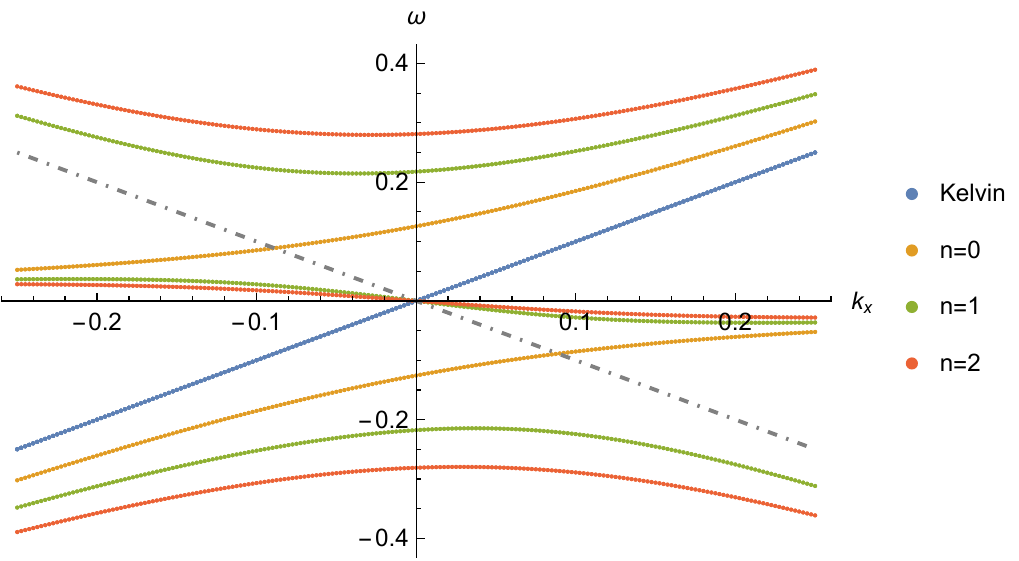}
\caption{Spectrum for ordinary equatorial hydrodynamics as originally derived in Ref.~\cite{Matsuno1966}. We took $\mathcal{G} = 0.001$. There are two chiral waves, which are the Kelvin wave and the chiral low-frequency part of the $n=0$ solution, which is called the Yanai wave.}    \label{fig:matsuno}
\end{figure}

 It is instructive to compare the details of this spectrum with that of ocean dynamics \cite{Matsuno1966} in which magnetic fields vanish as shown in Fig.~\ref{fig:matsuno}.
The spectrum of Fig.~\ref{fig:matsuno} has qualitative differences with respect to the case of a toroidal magnetic field in Fig.~\ref{fig:nonuniform}. In particular, the Yanai wave (orange curve) in Fig.~\ref{fig:matsuno} is a single continuous curve propagating westwards while when turning on toroidal magnetic fields it splits into two waves, namely, the slow and fast magneto-Yanai waves. The other qualitative feature is that the Yanai-wave and the Rossby-wave (green curve passing through $\omega=0$) in Fig.~\ref{fig:matsuno} satisfy the normalizability condition \eqref{eq:stabilityconstr} for all values of $k_x$. This means that contrary to the magneto-Yanai and magneto-Rossby waves in Fig.~\ref{fig:nonuniform}, the Yanai and Rossby waves are continuous and well defined as $k_x\to-\infty$. The remaining higher order modes are qualitatively similar in the case of toroidal or vanishing magnetic fields. 

We note that the spectrum of ocean dynamics presented in Fig.~\ref{fig:matsuno} is qualitatively similar to the spectrum of equatorial MHD waves in the presence of a constant magnetic field. We discuss this case in detail in Appendix \ref{app:constantB}. In the next section we study further the properties of equatorial MHD waves in the presence of toroidal magnetic fields using the f-plane approximation and show that magneto-Kelvin and magneto-Yanai waves are topologically protected and hence expected to be a robust feature of the equatorial MHD spectrum.

\section{Topology of plasma oscillations}
\label{sec:topology}
In the previous section, we obtained the equatorial MHD spectrum with toroidal magnetic fields. In this section we wish to understand what are the robust properties of this spectrum, that is, what properties are topologically protected, say by slight deformations of the Hamiltonian or by changes of the shape of the "interface" (equator) separating the north and south hemispheres. This analysis can be carried out by viewing each hemisphere as a distinct topological system separated by the equator that acts as an interface between the two "bulk" systems, as shown in \cite{Delplace1075} for ocean dynamics (see also \cite{green2020topological} for general axisymmetric surfaces). Here we generalise this analysis to MHD.

To begin with we define the "bulk" topological systems as the two systems governed by a Hamiltonian $\mathcal{H}$ at a point $\theta=\theta_0$ on each hemisphere away from the poles of the sphere and away from the equator. This approximation is what is commonly referred to as the f-plane approximation \cite{Gill1982-ks}. Within this approximation the system is translational invariant in the two spatial directions. Indeed, by looking at linear fluctuations around the equilibrium state detailed in Section \eqref{sec:MHDSW} using plane-waves of the form $\sim \exp(i \omega t - i \vec{k}.\vec x)$ we are able to write Eq.~\eqref{matrixrescaled11112} as
\begin{subequations} \label{eq:combin111}
       \begin{align}
         \left(   \omega  -   H   \right)   \begin{bmatrix}
      \delta \hat{h} &
          \delta u^{x }&
     \delta u^{y } &
   \delta   B^{x } &
    \delta  B^{y }
    \end{bmatrix}^{T} =0  ~~ , 
\end{align}
with 
 \begin{align}  \label{matrixresca12}
     H  =   \begin{bmatrix}
       0 &   k_x   &  k_y  &  0 & 0 \\  
          k_x   & 0  &   -i  m    &     k_x    m \gamma^2   &    0 
 \\ 
      k_y    &  i    m  &   0   &   0  &  k_x m  \gamma^2     \\ 
             0 &  k_x  m    & 0   &   0  & 0  \\       
                        0 & 0 & k_x m   & 0 & 0 
    \end{bmatrix}   ~~  , 
 \end{align} 
\end{subequations}
where $k^2 = k_x^2 + k_y^2$. For simplicity, we have assumed that we are far enough from the poles so that $\sin(\theta ) \approx 1 $ and $m\approx \pm 1$ with $m=+1$ for the northern hemisphere and $m=-1$ for the southern hemisphere. We note that because we are focusing on a specific point $\theta=\theta_0$ it does not matter whether the magnetic field is toroidal or constant in equilibrium. By solving Eq.~\eqref{eq:combin111} we find five modes. 
\begin{figure}
    \centering
    \includegraphics[width=8cm]{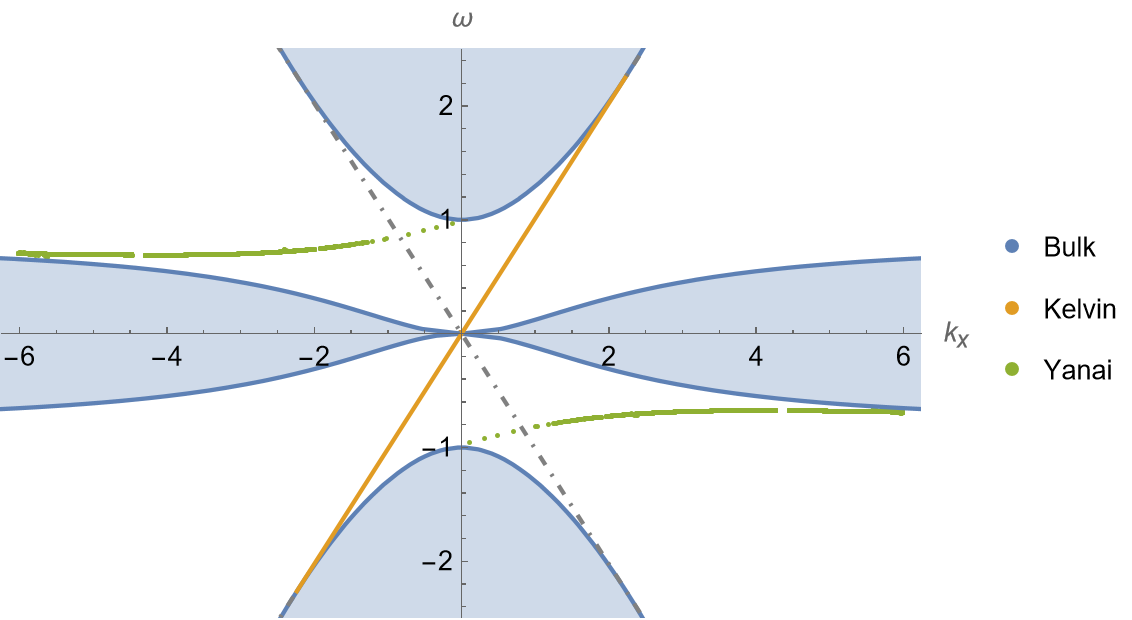}
\caption{f-plane spectrum including bulk bands (in blue) and interface modes (in orange and green) as derived in App.~\ref{app:berrycurvature}. The blue bands include the magneto-Rossby modes passing through $\omega=0$ when $k_x=0$ and the magneto-inertial-gravity modes with $\omega\ne0$. The orange curve is the chiral magneto-Kelvin wave and the green solid curve the chiral magneto-Yanai wave. The grey dashed line is the non-normalizable mode. We took $\gamma= 0.156$ and regulator $\epsilon = 0.2$ (see Appendix \ref{app:berrycurvature}).}    
\label{feiuw213213}
\end{figure}
The spectrum reveals the existence of a trivial mode with $\omega=0$, two magneto-Rossby modes that pass through $\omega=0$ for $k_x=0$ and two magneto-inertial-gravity modes that pass through $k_x=0$ for $\omega\ne0$. These modes correspond to the blue regions presented in Fig.~\ref{feiuw213213} for arbitrary $k_y$ and were derived in \cite{2007A&A...470..815Z}. Referring to each of these modes as "bands", of particular importance is the fact that there is a "band gap" in the spectrum with $m$ being the mass gap, that is, there is a finite distance in momentum space between the magneto-Rossby bands and the magneto-inertial-gravity bands where the two bands do not cross. This gap only closes for large values of $\gamma$. From the point of view of the bulk-interface correspondence in condensed matter \cite{KANE20133} this suggests the existence of topological properties in the spectrum. In particular, to each band we may associate a Chern number $\mathcal{C}$ whose difference across the two sides of the equator ($\Delta \mathcal{C}$) can reveal the number of localised chiral modes propagating at the equator. In order to extract the Chern number we must compute the Berry curvature $F_{xy}$, which for continuous systems typically requires introducing an ultra-violet, short-distance, cut-off. We discuss how to properly introduce this cut-off as well as other relevant technical details in Appendix \ref{app:berrycurvature}. The Chern number is then defined by appropriately integrating over the difference in Berry curvatures $\Delta F_{xy}$ between the two hemispheres for a single band. Choosing the band corresponding to the magneto-inertial-gravity waves we find
\begin{align} \label{eq:Cherncomputation}
\Delta  \mathcal{C} = \frac{1}{2 \pi    } \int d k_x d k_y \Delta   F_{xy } =2 ~~. 
\end{align}
This is the same result that is obtained for ocean dynamics with vanishing magnetic fields \cite{Delplace1075} and suggests that there are two topologically protected chiral modes propagating along the equator. 

In order to precisely identify the nature of these two chiral modes, we solve the equatorial MHD problem using a different approximation scheme. To wit, we consider applying the f-plane approximation to the case of a sharp interface by gluing the two hemispheres at $y=0$, as was done for ocean dynamics in \cite{Delplace1075}. This detailed setup is given in Appendix \ref{app:berrycurvature} and precisely leads to the two chiral modes presented in Fig.~\ref{feiuw213213}. The orange curve in Fig.~\ref{feiuw213213} is the magneto-Kelvin wave connecting the magneto-Rossby band to the magneto-inertial-gravity band for $k_x>0$ in the upper half plane. In turn the green curve in Fig.~\ref{feiuw213213} is the magneto-Yanai wave connecting the magneto-Rossby band at $k_x<0$ to the magneto-inertial-gravity band for $\omega\ne0$ at $k_x=0$ in the upper half plane. The dashed line is the non-normalizable mode equivalent to the one found in Fig.~\ref{fig:nonuniform}. This analysis clearly shows the topological nature of the magneto-Kelvin and magneto-Yanai waves. We emphasize that the topological properties of the spectrum are a consequence of the bulk Hamiltonian \eqref{matrixresca12} with associated Chern number \eqref{eq:Cherncomputation}. As such, the existence of 2 chiral modes should appear in a variety of contexts, e.g. with different boundary conditions for which \eqref{matrixresca12} is the bulk Hamiltonian. Indeed, in Appendix \ref{sec:topologicalinsulators} we show that by solving a similar problem with two edge boundaries on an infinite strip leads to 2 localised edge modes on each of the two boundaries. 

As we noted above, the "bulk" analysis that we carried out is applicable in both the case of uniform and non-uniform magnetic fields. From this point of view, one expects that if the interface is slightly deformed away from a "sharp interface" as in the case of the $\beta$-plane approximation, the chiral edge modes (magneto-Kelvin and magneto-Yanai) remain topologically protected. Focusing first on the case of constant magnetic fields $B=B_0$, which we reviewed in Appendix \ref{app:constantB}, we clearly see the presence of magneto-Kelvin (blue curve) and magneto-Yanai (orange curve) waves in Fig.~\ref{fig:uniform}. A close analysis reveals that these two modes are the only two propagating chiral modes while the remaining higher-order excitations in Fig.~\ref{fig:uniform} can be interpreted as non-chiral "bulk" modes. Thus in this context we see that the bulk-edge correspondence holds. The situation is not as a clear in the case of a non-uniform magnetic field for which the MHD spectrum is presented in Fig.~\ref{fig:nonuniform}. In this context, there are also propagating magneto-Kelvin and magneto-Yanai waves but only the magneto-Kelvin wave is chiral and the magneto-Yanai wave has been split into two branches of lower and upper magneto-Yanai waves. This suggests that the bulk-edge correspondence does not hold for the non-uniform magnetic field case. However, we want to clarify that the splitting of the magneto-Yanai wave may be a feature of the truncation that we employed when using the $\beta$-plane approximation (see e.g. \eqref{eq:almostharmonic1} which was expanded up to order $y^4$) and mapping it to a QHO problem. Thus this splitting may not necessarily be a physical feature of the actual spectrum. Indeed, in earlier literature in which a different truncation was used, the magneto-Yanai wave appeared to be split into 4 parts (see Fig.~\ref{feiuw2213313}). It would be useful to clarify this by attempting to solve the problem analytically employing more accurate $\beta$-plane approximations \cite{DELLAR_2011} or by
performing full numerical simulations as in \cite{Matilsky_2022, blume2023inertial} and extract the exact spectrum of equatorial MHD waves with non-uniform magnetic fields. It is our expectation that, given the topological properties of magneto-Kelvin and magneto-Yanai waves, such gap in the magneto-Yanai wave is actually not present in a full numerical simulation.

\section{Discussion}
\label{sec:discussion}
In this work we derived the spectrum of MHD oscillations focusing on the solar tachocline with toroidal magnetic fields. We first derived the magnetohydrodynamic shallow water equations imposing adjective boundary conditions and found the equations that coincide with those previously formulated by Gilman \cite{Gilman_2000}. We furthermore showed that the shallow water analogue of Gauss law becomes a redundant equation as is the case for ordinary MHD. By systematically expanding the magnetohydrodynamic equations which are subjected to the $\beta$-plane approximation near the equator, we uncovered new qualitative features of the solar tachocline spectrum. Particularly, we did not find slow magneto-Rossby waves as discussed in \cite{Zaqarashvili_2018}, which were previously argued to produce the 100-year period Gleissberg cycle \cite{Zaqarashvili_2018}. Assuming that we have solar waves with wave vector $k_x \sim \sqrt{\mathcal{G}}/2 $, the slowest magneto-Rossby wave corresponding to Fig.~\eqref{fig:nonuniform} has a period of around seven years. The second slowest wave appears due to the splitting of the magneto-Yanai wave caused by the non-uniform magnetic field. For $k_x \sim \sqrt{\mathcal{G}}/2 $, this wave has a period of around three years while the magneto-Kelvin wave has a period of around 2 years. These types of oscillations can potentially be correlated with solar annual oscillations and solar cycles. These results hold qualitatively for any star with more than 30\% of solar mass in which a tachocline layer is expected to be formed as long as $\mathcal{G}$ is small.

In the second part of this work we focused on understanding topological poperties of MHD plasma oscillations. In order to do so, we studied the MHD shallow water dynamics from the point of view of band theory and found topological properties associated with the MHD shallow water Hamiltonian in the "bulk" (i.e. away from the equator and the poles). This allowed us to associate a Chern number to the upper band (i.e. magneto-inertial-gravity waves) whose difference across the equator yields a topological invariant: the total number of chiral edge modes. By explicitly solving for the equatorial spectrum using the f-plane approximation we deduced that these two chiral modes correspond to the magneto-Kelvin and magneto-Yanai waves. These two modes are robust properties of the equatorial MHD spectrum as they are stable against deformations of the Hamiltonian or deformations of the equator. This motivates further taking into consideration the oscillations caused by magneto-Kelvin and magneto-Yanai waves as potential causal explanations for different types of solar activity.

The methods employed here can in principle be used to study a variety of different contexts, such as axisymmetric geometries as in \cite{green2020topological}, the inclusion of dissipative effects such as viscosity and resistivity in MHD \cite{Armas:2018atq, Armas:2018zbe, Armas:2022wvb}, different magnetic field configurations such as double band magnetic field configurations \cite{Belucz_2023}, oblateness effects \cite{vanBaal:2020imd}, as well as the topological properties of the magnetic bouyancy instability as in \cite{Leclerc_2022, leclerc2023exceptional}. In a related direction, it would be interesting to study whether topology plays a role when considering a stratified (compressible) fluid structure within the Sun in which case novel effects appear such as thermal Rossby waves and retrograde vorticity modes \cite{blume2023inertial}. We leave these interesting directions for future research. \\

\paragraph*{Acknowledgements}
We thank Gerrit M. Horstmann, Teimuraz Zaqarashvili, Clement Tauber and Jasper van Wezel for useful discussions. The work of JA is partly supported by the Dutch Institute for Emergent Phenomena (DIEP) cluster at the University of Amsterdam via the programme Foundations and Applications of Emergence (FAEME).

\onecolumngrid

\appendix
\section{Derivation of the shallow water magnetohydrodynamics equations}
\label{app:shallowaterequations}
In order to derive the shallow water MHD equations we consider a three dimensional magnetohydrodynamics fluid placed in between a (top) dynamical interface located at $F=Z-\eta(t,X,Y)=0$ and a (bottom) rigid boundary located at $G=Z+H(X,Y)=0$ along the $Z$ direction, $\eta$ is the field that accounts from changes in height due to a dynamical interface while $H$ is the topography. Here $X,Y$ are spatial directions and $t$ the time direction. In the solar context, the rigid boundary can model the boundary between the radiative zone and the bottom of the tachocline while the top interface models the boundary between the radiative and the overshoot part of the tachocline. Alternatively, the rigid boundary can model the bottom of the overshoot part of the tachocline while the interface models the boundary between the overshoot part of the tachocline and the convective zone (see e.g. \cite{Gilman_2000}). The magnetohydrodynamics equations governing the three dimensional incompressible bulk fluid are given by 
\begin{equation} \label{eq:3dMHD}
\begin{split}
    \partial_t\rho+\tilde\nabla_A(\rho v^A) &=0~~, \\ 
    \tilde \nabla_A v^A  &=0~~,\\
    \rho \tilde D_t v^A&=-\tilde \nabla^A\left(P+\frac{1}{2\mu_0}B^2\right)+\frac{1}{\mu_0}B^B\nabla_B B^A+\rho g \delta^A_Z~~,\\
    \partial_t B^A&=B^B\tilde \nabla_B v^A - v^B\tilde \nabla_B B^A~~, \\
    \tilde\nabla_A B^A&=0~~,
\end{split}
\end{equation}
where $\tilde \nabla$ is the three dimensional covariant derivative associated with the three dimensional spatial metric $g_{AB}$, the operator $\tilde D_t$ is defined as $\tilde D_t=\partial_t +v^A \tilde \nabla_A$, $\rho$ is the mass density of the three dimensional fluid velocity, $P$ is the pressure, $B^A$ the three dimensional magnetic field, $\mu_0$ the magnetic permittivity and $g$ the acceleration of gravity. The indices $A,B,...$ run over the three spatial directions $X,Y,Z$. We parameterise the components of $v^A$ and $B^A$ according to $v^A=(u^\mu, v^Z)$ and $B^A=(B^\mu, B^Z)$ where the indices $\mu,\nu...$ run over the directions $X,Y$. We note that the fact that the fluid is incompressible means that the density $\rho$ is constant, i.e. $\tilde D_t \rho=0$. For the solar tachocline this is justified given that the thickness of the tachocline is much smaller than the length scale of density variations across the tachocline \cite{Gilman_2000}. Also note that the last equation in \eqref{eq:3dMHD}, the Gauss law, is a constraint equation on an initial Cauchy slice since $\tilde D_t \tilde \nabla_A B^A=0$ and hence only needs to be satisfied for an initial magnetic field configuration. Eqs.~\eqref{eq:3dMHD} must be supplemented with boundary conditions for the fluid velocity and magnetic field at the bottom boundary and the interface. We take advective boundary conditions for the fluid velocity and similarly for the magnetic field as expected for the solar tachocline \cite{Gilman_2000},
\begin{equation} \label{eq:3dboundary}
\begin{split}
\tilde D_t F|_{Z=\eta}=0~~,~~B^A \tilde \nabla_A F|_{Z=\eta}=0~~,\\
\tilde D_t G|_{Z=-H}=0~~, B^A \tilde  \nabla_A G|_{Z=-H}=0~~.\\
\end{split}
\end{equation}
We also need to specify boundary conditions for the pressure $P$ and the modulus of the magnetic field $B^2$ at $Z=\eta$, in particular
\begin{equation} \label{eq:3dboundary2}
    P|_{Z=\eta}=P_0~~,~~B^2|_{Z=\eta}=B^2_0~~,
\end{equation}
for constant $P_0$ and $B^2_0$. To make further progress we focus on geometries which are trivial along the $Z$ direction and hence focus on metrics that take the form 
\begin{equation}
g_{AB}dX^AdX^B=dZ^2+g_{\mu\nu}dX^\mu dX^\nu~~,    
\end{equation}
where $g_{\mu\nu}$ is an arbitrary two dimensional spatial metric independent of $Z$. In the main text we take $g_{\mu\nu}$ to be the two dimensional spherical metric but here, for completeness, we leave it arbitrary. Under this assumption the boundary conditions \eqref{eq:3dboundary} yield
\begin{equation} \label{eq:3bdoundary3}
\begin{split}
v^Z-D_t \eta|_{Z=\eta}=0~~,~~B^{Z}-B^{\mu}\nabla_\mu \eta|_{Z=\eta}=0~~,\\
v^Z+u^\mu\nabla_\mu H|_{Z=-H}=0~~,~~B^{Z}+B^{\mu}\nabla_\mu H|_{Z=-H}=0~~,
\end{split}    
\end{equation}
where the operator $D_t$ is defined as $D_t=\partial_t +u^\mu\nabla_\mu$ with $\nabla_\mu$ being the covariant derivative associated to the spatial metric $g_{\mu\nu}$. In order to proceed further we must specify a gradient ordering for the hydrodynamic expansion. We take $\eta\sim\mathcal{O}(1)$ in order to account for interface effects at the same order as the fluid velocity $u^\mu\sim\mathcal{O}(1)$ \footnote{Note that the scaling $\eta\sim\mathcal{O}(1)$ is different than the scaling $\eta\sim\mathcal{O}(\partial^{-1})$ considered in the context of droplets \cite{Armas:2015ssd, Armas:2016xxg, Armas:2018ibg}.}. In this case, the boundary conditions \eqref{eq:3bdoundary3} tell us that $v^Z\sim B^{Z}\sim\mathcal{O}(\partial)$ given that we take $\eta\sim\mathcal{O}(1)$. This means that we can typically ignore terms involving $v^Z$ and $B^Z$. For instance, if we pick the $Z$ component of the third equation in \eqref{eq:3dMHD} we can ignore $v^Z$ and $B^Z$ terms. Integrating it from $Z$ to $\eta$ we obtain the pressure 
\begin{equation} \label{eq:pressureequation}
P+\frac{1}{2\mu_0}B^2=P_0+\frac{1}{2\mu_0}B^2_0+\rho g (\eta-Z) + \mathcal{O}(\partial)~~,   
\end{equation}
where we have used \eqref{eq:3dboundary2}. Integrating the remaining equations in \eqref{eq:3dMHD} from $Z=-H$ to $Z=\eta$ and using \eqref{eq:3bdoundary3} and \eqref{eq:pressureequation}, we obtain
\begin{equation} \label{eq:2dshallow}
\begin{split}
\partial_t \eta +\nabla_\mu\left((\eta+H)u^\mu\right)=0~~,\\
D_t u^\mu=g\nabla^\mu \eta+\frac{1}{\mu_0\rho}B^\nu\nabla_\nu B^\mu~~,\\
D_t B^\mu=B^\nu\nabla_\nu B^\mu~~,\\
\nabla_\mu\left((\eta+H)B^\mu\right)=0~~.
\end{split}    
\end{equation}
By defining the height $h=\eta+H$ we straightforwardly obtain Eqs.~\eqref{eq:allequations}. We note that the $Z$ component of the fourth equation in \eqref{eq:3dMHD} does not feature in \eqref{eq:2dshallow} since it is of order $\mathcal{O}(\partial^2)$ and hence can be neglected.

\section{Comparison with earlier literature}
\label{app:zaqa}
\begin{figure} [h!]
    \centering
    \includegraphics[width=8cm]{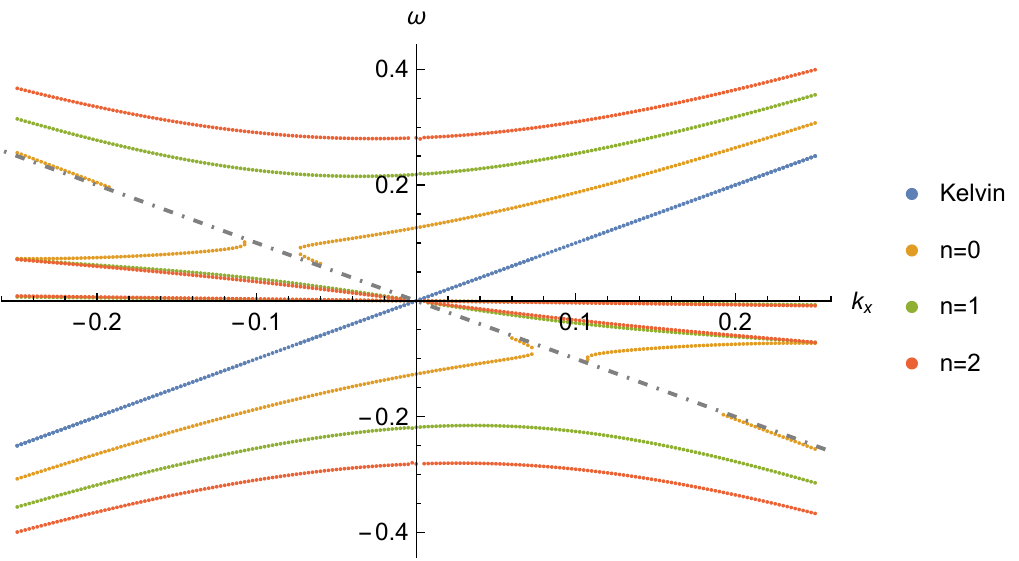}
\caption{Spectrum found in Ref.~\cite{Zaqarashvili_2018} by ignoring the coefficient $D$ with $\gamma= 1.56$ and $\mathcal{G} = 0.001$. The blue curve is the magneto-Kelvin wave, the orange lines are the magneto-Yanai waves, the remaining lines contain a set of 2 magneto-Rossby waves and one magneto-inertial-gravity wave for each $n\ge1$. The dashed line is the (unphysical) non-normalizable mode with $\omega=-k_x$.}    \label{feiuw2213313}
\end{figure}

As we mentioned in the main text, the spectrum of equatorial MHD waves in the solar tachocline had previously been derived in Ref.~\cite{Zaqarashvili_2018}. However, the spectrum derived in \cite{Zaqarashvili_2018} differs from the one we obtained in Fig.~\ref{fig:nonuniform} in several ways. These differences are rooted in the fact that Ref.~\cite{Zaqarashvili_2018} has arrived at an equation analogous to Eq.~\eqref{eq:almostharmonic} which takes the form 
\begin{align}  \label{eq:almostharmoniczaqa}
     \partial_y^2  \delta u^y   =  \left[ A y^2   - B   + \mathcal{O} (y^4)  \right]   \delta u^y  +  2  y  \left[   C   + \mathcal{O} (y^2)    \right]  \partial_y   \delta u^y  ~~  . 
      \end{align}
Comparison between Eq.~\eqref{eq:almostharmoniczaqa} and Eq.~\eqref{eq:almostharmonic} reveals that the spectrum obtained in Fig.~\ref{feiuw2213313} did not consider the existence of the coefficient $D$ appearing in \eqref{eq:ABCDcoef}. However, as explained in the main text, the coefficient $D$ is required for the consistency of the quantum harmonic oscillator equation \eqref{eq:finalequation}. Indeed, if we had ignored the coefficient $D$ and computed the spectrum using Eq.~\eqref{eq:almostharmoniczaqa} we would find the spectrum depicted in Fig.~\ref{feiuw2213313}, as was obtained in Ref.~\cite{Zaqarashvili_2018}.
Comparing the spectrum in Fig.~\ref{feiuw2213313} with that of Fig.~\ref{fig:nonuniform} we see various qualitatively differences. In particular, the spectrum of Fig.~\ref{feiuw2213313} contains another set of modes for each $n\ge1$ that are referred to as "slow magneto-Rossby waves" in Ref.~\cite{Zaqarashvili_2018} which "hover" slightly above $\omega=0$. These modes are absent from Fig.~\ref{fig:nonuniform} when including the coefficient $D$. In addition, we see that in Fig.~\ref{feiuw2213313} the magneto-Yanai wave (in orange) apears to be discontinuous and split into various parts around the non-normalisable solution $\omega=-k_x$. This behaviour is absent in the spectrum of Fig.~\ref{fig:nonuniform} in which the magneto-Yanai wave is composed of two continuous curves. Furthermore, in Fig.~\ref{feiuw2213313} the lower part of the magneto-Yanai wave touches the magneto-Rossby wave at a finite value of $k_x<0$ in the upper half plane. In contrast, the magneto-Yanai wave does not touch the magneto-Rossby wave in Fig.~\ref{fig:nonuniform} due to the normalization condition in Eq.~\eqref{eq:stabilityconstr}.

\section{Equatorial spectrum with constant magnetic fields}
\label{app:constantB}
\begin{figure}[h!]
\centering
	\subfloat{ \includegraphics[width=8cm]{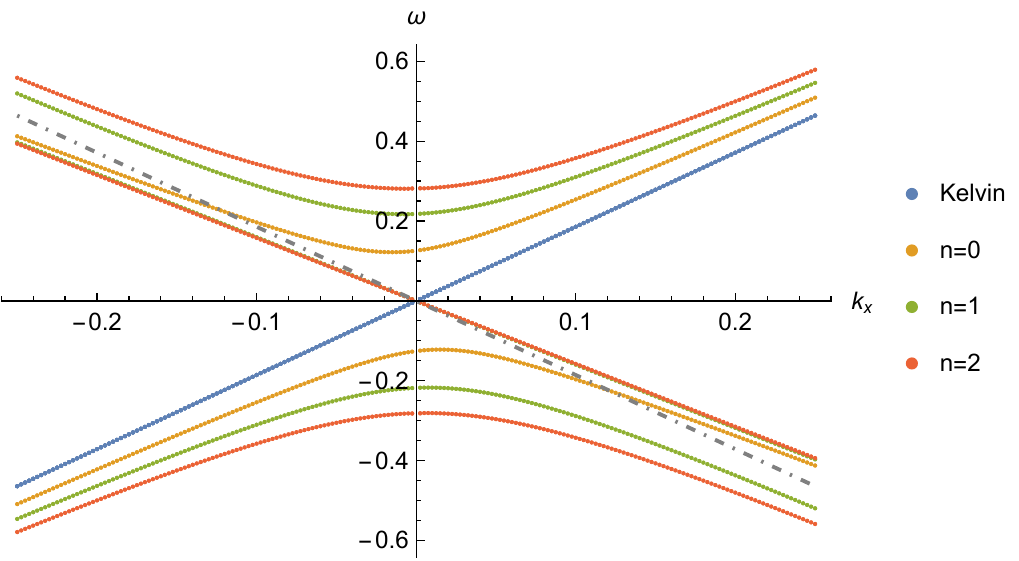}}
	\subfloat{\includegraphics[width=8cm]{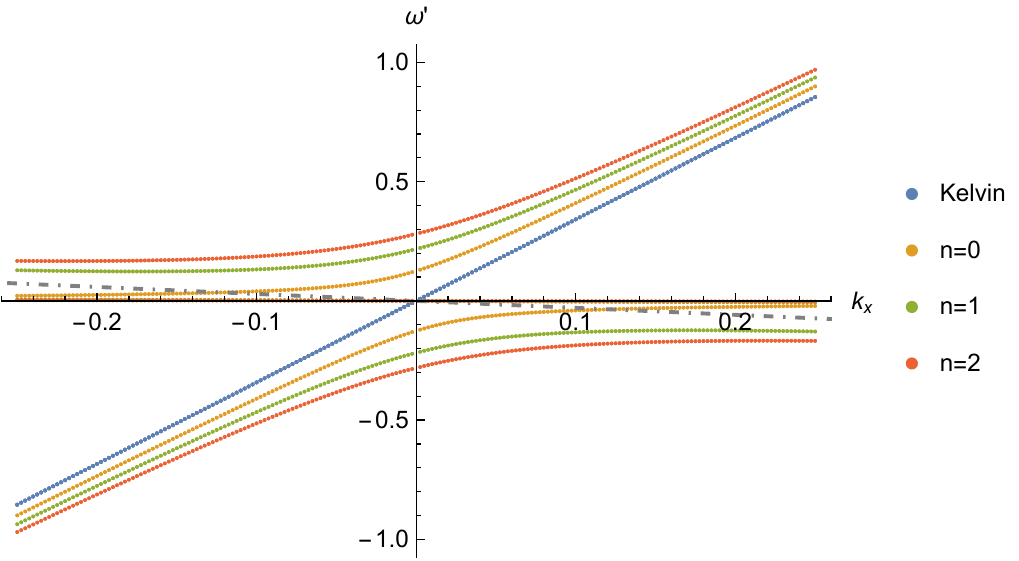}}
\caption{Spectrum for equatorial magnetohydrodynamics with a uniform magnetic field. The figure on the right hand side is the spectrum as computed earlier in Ref.~\cite{Zaqarashvili_2018}. The figure on the right side is the same spectrum in the Alfv\'{e}n frame $\omega \to \omega'=\omega  +\gamma k$. We took the values $\gamma= 1.56$ and $\mathcal{G} = 0.001$.}
    \label{fig:uniform}
\end{figure}

Even though not being the most relevant situation for the solar tachocline, it is interesting to consider the case in which the equilibrium configuration has a uniform magnetic field, i.e. $B^{\mu}_0 = \delta^{\mu}_{\phi } B_0$. The spectrum of equatorial magnetohydrodynamics in this case was derived in Ref.~\cite{Zaqarashvili_2018} and here we briefly review this spectrum but give it a slightly different perspective. Following the procedure of Section \ref{sec:MHDSW} for this equilibrium configuration we obtain the magneto-Kelvin solution with dispersion relation $\omega = \sqrt{1+\gamma^2} k $. The remaining solutions can be obtained again by mapping the differential equations to a QHO equation. In this case a transformation like Eq.~\eqref{eq:transformation} is not needed and one is led to the spectrum given in Fig.~\ref{fig:uniform} on the left hand side.
At first sight the spectrum appears to be somewhat different than the case of vanishing magnetic fields of Fig.~\ref{fig:matsuno} but in fact the only qualitative difference is that the magneto-Rossby waves (the $n\ge1$ solutions passing through $\omega=0$) are tilted in the clockwise direction. In fact, instead of analysing the spectrum in a frame co-rotating with the fluid at the equator, we can adjust the boost to a frame co-rotating with the Alfv\'{e}n wave by performing the transformation $\omega \to \omega'=\omega  +\gamma k$ leading to the right hand side of Fig.~\ref{fig:uniform}. In this Alfv\'{e}n frame of reference, we can clearly see the existence of two chiral modes, the magneto-Kelvin (blue curve) and magneto-Yanai (orange curve) waves which connect the magneto-Rossby waves to the magneto-inertial-gravity waves as in Fig.~\ref{fig:matsuno} for the case of vanishing magnetic fields. The remaining curves constitute the higher-order modes for $n\ge1$ while the dashed gray line is a non-normalisable mode. The topological analysis we performed in Section \ref{sec:topology} also applies to the spectrum of Fig.~\ref{fig:uniform} and indeed we again find the existence of two chiral modes consistent with the bulk-edge correspondence.

\section{f-plane, Berry curvature and topological insulators}
\label{app:berrycurvature}

In Section \ref{sec:topology} we introduced the f-plane approximation both away from the equator and at the equator. We also discussed the computation of the Berry curvature. In this section we give further details on these aspects and also consider the analogue setup of a "topological insulator" to highlight the topological origin of the magneto-Kelvin and magneto-Yanai waves. 

\subsection{f-plane approximation and Berry curvature}
\begin{figure} [h!]
    \centering
    \includegraphics[width=8cm]{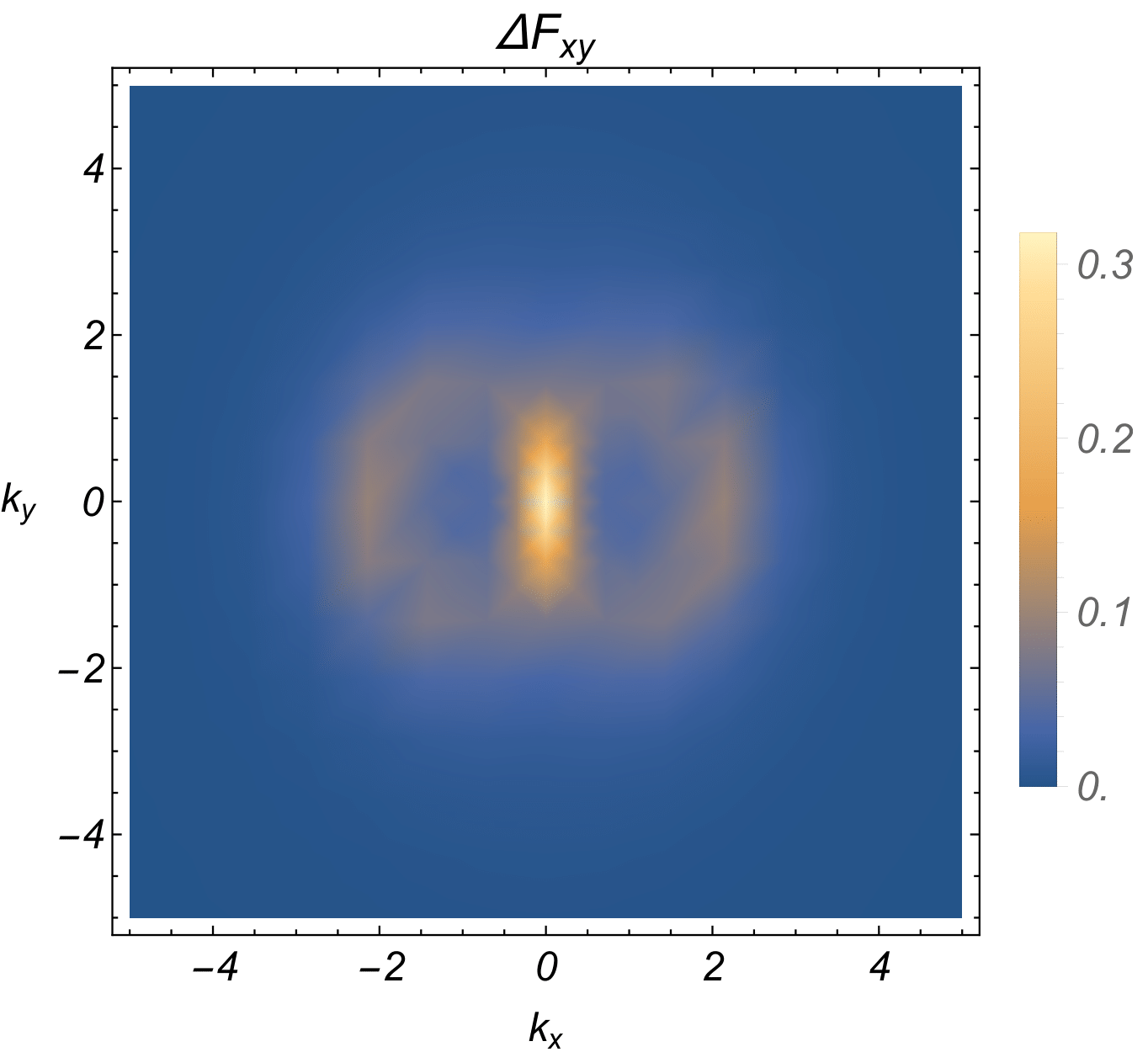}
\caption{Numerical result for the Berry curvature difference $\Delta F_{xy}$ as a function of $k_x$ and $k_y$. We took $\epsilon=0.2$.}
\label{berrycurvatureeeee}
\end{figure}

As discussed in Section \ref{sec:topology}, the calculation of the Berry curvature for continuous systems and, in particular, of the equatorial spectrum using the f-plane approximation requires the introduction of a ultra-violet, short-distance, cut-off. It was shown in \cite{souslov2019topological, delplacereal} that the introduction of a higher-order gradient correction in the system of Eqs.~\eqref{eq:allequations}, namely odd viscosity, provides a natural regulator for continuous systems with shallow water-like dynamics. We will adopt this regularization procedure here but we note that other regularization schemes are possible and we will discuss them in a future publication. Odd viscosity can be introduced by modifying the momentum dynamics given by the second equation in \eqref{eq:allequations} to
\begin{align}
    \label{eq:viscosityyyy1}
 D_t u^{\mu}    & = \nu_o \varepsilon^{\mu \nu } 
\nabla^2  u_{\nu} - g  \nabla^{\mu}  h + \frac{1}{\mu_0 \rho} B^{\nu} \nabla_{\nu} B^{\mu}  ~~ , 
\end{align}
where $\nu_0$ is the (constant) odd viscosity coefficient appearing in front of a term that is second order in gradients and $\epsilon^{\mu\nu}$ is the two-dimensional Levi-Civita tensor. Taking into account this modification, we can extract the Hamiltonian by performing perturbations around an equilibrium state with constant or toroidal magnetic fields to obtain an equation of the form \eqref{eq:combin111} but with Hamiltonian given by
 \begin{align}  \label{matrixresca124}
   \mathcal{H}  =   \begin{bmatrix}
       0 &   k_x   &  k_y  &  0 & 0 \\  
          k_x   & 0  &   -i  (m - \epsilon k^2 )    &     k_x    m \gamma^2   &    0 
 \\ 
      k_y    &  i    (m - \epsilon k^2 )  &   0   &   0  &  k_x m  \gamma^2     \\ 
             0 &  k_x  m    & 0   &   0  & 0  \\       
                        0 & 0 & k_x m   & 0 & 0 
    \end{bmatrix}   ~~  , 
 \end{align} 
where $m = \pm 1$ depending on whether one is in the upper or lower hemisphere and where we have defined $\epsilon \equiv \frac{2 \nu_o \Omega_0 }{g h_0}$. We note that the Hamiltonian \eqref{matrixresca124} reduces to that of \eqref{matrixresca12} when $\nu_0=0$. The dispersion relations (eigenvalues) that this leads to by means of \eqref{eq:combin111} are depicted in blue in Fig.~\ref{feiuw213213}. We now wish to compute the Berry curvature associated to the eigenvalues, in particular to the upper "bands", that is the magneto-inertial-gravity waves. Typically this can be done by extracting the eigenvectors of the Hamiltonian. However, because we are dealing with a 5 x 5 matrix this is analytically difficult. Instead the Berry curvature $F^{\pm }_{xy}$ with the $\pm$ signs indicating the upper (+) magneto-inertial-gravity band and the lower (-) magneto-Rossby band, can be extracted directly by looking at the Hamiltonian and evaluating 
\begin{align} \label{eq:Berry}
  F^{\pm }_{xy }  =  \frac{1}{2}  \text{Res}  \left( \text{tr} [ G \partial_{k_x } \mathcal{H} G^2 \partial_{k_y } \mathcal{H}    ] - \text{tr} [ G \partial_{k_y } \mathcal{H} G^2 \partial_{k_x } \mathcal{H}    ] ,  W = \omega_{\pm }
 \right) ~~ , 
\end{align}
with $G = 1 / ( W - \mathcal{H} )$ \cite{Kapustin_2020} and where $\omega_\pm$ denotes the solution for the dispersion relation for the upper and lower bands, respectively. In turn, the difference between Berry curvatures in two hemispheres is given by
\begin{align} \label{eq:diffBerry}
          \Delta F_{xy} =  F^{\pm}_{xy} \big|_{m=1}  - F^{\pm}_{xy} \big|_{m= - 1} ~~ .       
\end{align}
$  \Delta F_{xy}$ is computed numerically and given in Fig.~\ref{berrycurvatureeeee} as a function of $k_x$ and $k_y$. 
Using the Hamiltonian \eqref{matrixresca124} in \eqref{eq:Berry}, focusing on the upper (+) band, and numerically integrating \eqref{eq:diffBerry} over $k_x$ and $k_y$ leads to the result for the Chern number as presented in \eqref{eq:Cherncomputation}. We note that for the particular case of the Hamiltonian \eqref{matrixresca124} it is possible to verify that the Berry curvature $F_{xy}$ vanishes for large $k_x$ and $k_y$ implying that it was not necessary to introduce a regulator in this case. However, the regulator is needed when solving for the equatorial spectrum in the f-plane approximation as will be done below.

\subsection{Equatorial spectrum in the f-plane approximation}
Here we show how to compute the f-plane spectrum of Fig.~\ref{feiuw213213} by gluing together solutions on the upper and lower hemispheres \cite{delplacereal}. These solutions are necessariyl non-uniform along the $y$-direction, and hence we should consider an ansatz of the form $\sim \exp(i\omega t -i k_x x  )$ for the plane wave perturbations. Including the presence of the odd-viscosity the equations of motion \eqref{eq:allequations} lead to
  \begin{align}
        \label{matrixresc09812}
\begin{bmatrix}
      - \omega  &   k_x   &  i \partial_y  &  0 & 0 \\  
          k_x   &   - \omega  &   -i  (m + \epsilon  ( \partial_y^2   -k_x^2  )  )    &     k_x     \gamma^2   &    0 
 \\ 
      i \partial_y    &  i    (m + \epsilon  ( \partial_y^2   -k_x^2  )  )  &     - \omega   &   0  &  k_x   \gamma^2     \\ 
             0 &  k_x      & 0   &     - \omega & 0  \\       
                        0 & 0 & k_x    & 0 &   - \omega 
    \end{bmatrix}    \begin{bmatrix}
      \delta \hat h   \\ 
          \delta u^{x } \\ 
     \delta u^{y  } \\ 
   \delta   B^{x } \\ 
    \delta  B^{ y  }
    \end{bmatrix}   =0   ~~  , 
 \end{align} 
with $m$ changing sign accross the equator. We proceed by solving Eq.~\eqref{matrixresc09812} for $\delta B^x$, $\delta B^y$ and $\delta \hat h$ in order to find the pair of equations
 \begin{subequations} \label{eq:dedaluseqs}
 \begin{align}
    i  \left( -\epsilon ( \partial_y^2 
 - k_x^2)+\frac{k_x \partial_y}{\omega } -m \right) \delta u^y & =  \frac{\omega^2 - \left(\gamma ^2+1\right) k_x^2}{\omega }  \delta u^x ~~ ,  \\ 
   i  \left(\frac{k_x \partial_y}{\omega }+\epsilon ( \partial_y^2 - k_x^2 )+m\right) \delta u^x &  =   \frac{ \omega ^2 -\gamma ^2 k_x^2+\partial_y^2 } {\omega } \delta u^y ~~ . 
 \end{align}   
 \end{subequations}
There are different types of solutions that can be extracted from here. \paragraph{magneto-Kelvin wave.} The first type of solutions is the magneto-Kelvin wave satisfying $\omega^2 = \left(1 + \gamma ^2\right) k_x^2$. This leads to $\delta u^y =0 $ and $\delta u^x \sim e^{q y }$ for some $q$ where $q$ is given by a solution to the equation
\begin{align}
    \frac{k_x q }{\omega }+\epsilon ( q^2  - k_x^2 )+m   =0  ~~ .
\end{align}
Explicitly solving the equation above leads to two possibilities
\begin{equation} \label{eq:q12}
     q^{(1,2)} = -\frac{k}{\omega } \pm \frac{  \sqrt{4   \epsilon  \left(k^2   \epsilon -m  \right)+ \frac{k^2}{\omega^2 }}}{2   \epsilon }  ~~  . 
 \end{equation}
We first consider the case in which $\omega = +\sqrt{1 + \gamma^2} k_x $. It holds that for $|k_x| < \sqrt{ \epsilon^{-1}}$, we have the following behaviour for $q^{(1,2)}$, namely
\begin{subequations}
\begin{align} 
      q^{(1)}  &  <   0 ~~ , ~~    q^{(2)}  <  0 ~~ , ~~ y > 0  ~~ .  \\ 
                   q^{(1)}    & >    0 ~~ , ~~      q^{(2)}   <   0 ~~ , ~~ y <  0  ~~ .
\end{align}    
\end{subequations}
We note that there are two bounded (physical) solutions in the upper hemisphere $y>0$ and one bounded solution in the lower hemisphere. Thus, in principle it should be possible to glue one solution from each hemisphere together. Specifically, identifying the upper hemisphere solutions with $\uparrow$ and lower hemisphere solutions with $\downarrow$, we have for $|k_x|< \sqrt{\epsilon^{-1}}$ that
 \begin{subequations}
      \begin{align}
     u_{\uparrow}  &  =  A_{(1)} \exp(q_{\uparrow}^{(1)} y ) +  A_{(2)} \exp(q_{\uparrow}^{(2)} y ) ~~ ,  \\ 
          u_{\downarrow}  &  =  B_{(1)} \exp(q_{\downarrow}^{(1)} y )  ~~ ,  
 \end{align}
 \end{subequations}
 where $A_{(1)}$ and $B_{(1)}$ are constant arbitrary coefficients. Imposing the gluing conditions \cite{delplacereal} in order to have a smooth solution across the equator we find
 \begin{align}
    \delta u_{\uparrow}^x \big|_{y =0 } = \delta u_{\downarrow}^x \big|_{y =0 } ~~ , ~~     \partial_y \delta u_{\uparrow}^x \big|_{y =0 }  =   \partial_y \delta u_{\downarrow}^x \big|_{y =0 }   ~~ , 
\end{align}
leads to
\begin{subequations}
      \begin{align}
     u_{\uparrow}  &  = A_{(1)} \left(\exp(q_{\uparrow}^{(1)} y ) + \frac{q^{(1)}_{\downarrow}-q^{(1)}_{\uparrow} }{q^{(2)}_{\uparrow}  -q^{(1)}_{\downarrow}} 
\exp(q_{\uparrow}^{(2)} y ) \right) ~~ ,  \\ 
          u_{\downarrow}  &  = A_{(1)}   \frac{ q^{(1)}_{\uparrow}-q^{(2)}_{\uparrow}}{q^{(1)}_{\downarrow}-q^{(2)}_{\uparrow}} \exp(q^{(1)}_{\downarrow} y )  ~~ . 
 \end{align}
 \end{subequations}
These conditions determine the magneto-Kelvin wave $\omega = \sqrt{1 + \gamma^2} k $ as a valid solution for $|k_x| < \sqrt{\epsilon^{-1}}$ corresponding to the orange line in Fig.~\ref{feiuw213213}.  
\paragraph{Unbounded solution.} For $\omega = -\sqrt{1 + \gamma^2} k_x $, we have for $|k_x| < \sqrt{ \epsilon^{-1}}$ that the pair of solutions in \eqref{eq:q12} behave as
 \begin{subequations}
\begin{align}
      q^{(1)} &  >   0 ~~ , ~~    q^{(2)}  >  0 ~~ , ~~ y > 0  ~~ .  \\ 
                   q^{(1)}    & <    0 ~~ , ~~      q^{(2)}  >  0 ~~ , ~~ y <  0  ~~ . 
\end{align}    
\end{subequations}
We see that there is no bounded solution in the upper hemisphere so this solution should be discarded as unphysical. This corresponds to the dashed line in Fig.~\ref{feiuw213213}.

\paragraph{magneto-Yanai wave.}
Now we consider $\omega^2 \neq (1 + \gamma^2 )k^2 $ and solve for $     \delta u^{x  }   $ to find
 \begin{align}
\begin{split}
     & \bigg[\gamma ^2 k_x^2  \left( ( 1 + \gamma ^2 )  k_x^2 -\partial_y ^2\right)      -\omega ^2 \left(k_x^2 \left(2 \gamma ^2 +1\right)+\left(\epsilon  \left(\partial_y ^2-k_x^2\right)+m\right)^2-\partial_y ^2\right)+\omega ^4 \bigg] \delta u^y  =0 ~~ .     
\end{split}
 \end{align}
Taking the ansatz $\delta u^y \sim \exp(s y )$, we can solve the above equation in order to find the four solutions
\begin{subequations} \label{eq:solutions}
    \begin{align} 
s^{(1,2)}_{ \pm  }  =  \pm  \sqrt{S_{(1,2)}} ~~ ,   
\end{align}
where $S_{(1,2)}$ is given by
\begin{align}
    S_{(1,2)}  =   \frac{\omega^2  - \gamma ^2 k^2 +2 \omega ^2 \epsilon  \left(k^2 \epsilon -m\right)  \pm \sqrt{\left(\omega ^2-\gamma ^2 k^2  \right) \left(\omega ^2 \left(-4 m \epsilon +4 \omega ^2 \epsilon ^2+1\right)-\gamma ^2 k^2  \left(4 \omega ^2 \epsilon ^2+1\right)\right)} }{2 \omega ^2 \epsilon ^2}  ~~ . 
\end{align}
\end{subequations}
This solution implies that a relation between $\delta u^y$ and $\delta u^x$, namely
\begin{subequations}
    \begin{align}
    \delta u^x   = \lambda \delta u^y  ~~ , 
\end{align}
where $\lambda$ is given by 
\begin{align}
   \lambda =  -\frac{i  \left(k_x^2 \omega  \epsilon +k_x s -\omega  \left(m+ s^2  \epsilon \right)\right)}{k_x^2 \left(\gamma ^2 +1\right)-\omega ^2} ~~ .  
\end{align}
\end{subequations}
There are $s_-$ solutions in Eq.~\eqref{eq:solutions} that are stable for $y> 0 $ and are identified with the upper hemisphere using $\uparrow$, and the $s_+$ solutions there are solutions stable for $y < 0$ and will be identified with the lower hemisphere $\downarrow$. This means that there are two bounded solutions on each side of the equator. To glue the solutions at the equator, and obtain smooth solutions we impose the gluing conditions \cite{delplacereal}
\begin{align} \label{eq:glue}
    \delta u^x_{\uparrow} \big|_{y =0 } =    \delta u_{\downarrow}^x \big|_{y =0 }  ~~ , ~~    \delta u_{\uparrow}^y \big|_{y =0 } = \delta u_{\downarrow}^y \big|_{y =0 }  ~~ , ~~   \partial_y \delta u_{\uparrow}^x  \big|_{y =0 } =    \partial_y \delta u_{\downarrow}^x \big|_{y =0 }  ~~ , ~~     \partial_y \delta u_{\uparrow}^y \big|_{y =0 }  =   \partial_y \delta u_{\downarrow}^y \big|_{y =0 }   ~~ , 
\end{align}
which amounts to the equations to solving the master equation
\begin{align} \label{eq:matrixsolveeee}
    \det  \begin{bmatrix}
          1 &  1 & -1 &  -1 \\ 
          \lambda^{(1)}_{\uparrow}  &     \lambda^{(2)}_{\uparrow} &     -\lambda^{(1)}_{\downarrow} &     - \lambda^{(2)}_{\downarrow}  \\
           s^{(1)}_{\uparrow} &  s^{(2)}_{\uparrow} & -s^{(1)}_{\downarrow} &  -s^{(2)}_{\downarrow} \\ 
        s^{(1)}_{\uparrow}     \lambda^{(1)}_{\uparrow}  &        s^{(2)}_{\uparrow} \lambda^{(2)}_{\uparrow} &       -s^{(1)}_{\downarrow} \lambda^{(1)}_{\downarrow} &     -  s^{(2)}_{\downarrow}  \lambda^{(2)}_{\downarrow}  \\
    \end{bmatrix}     =0  ~~ . 
\end{align}
This equation is hard to solve analytically but we can numerically compute the left-hand side of Eq.~\eqref{eq:matrixsolveeee} for different values of $\omega$ and $k_x$ and look for points where it goes below a certain numerically small threshold. Solving it yields the green lines in Fig.~\ref{feiuw213213}, which are the magneto-Yanai waves. Note that the Alfv\'{e}n dispersion relation $\omega_A = \pm \gamma k_x$ solves the master equation \eqref{eq:matrixsolveeee} but not does solve the gluing conditions \eqref{eq:glue} and therefore it is spurious and we discard it.

\subsection{Topological insulators}
\label{sec:topologicalinsulators}
\begin{figure*}[h!]
    \centering
	\subfloat[$\gamma=0$]{\includegraphics[width=8cm]{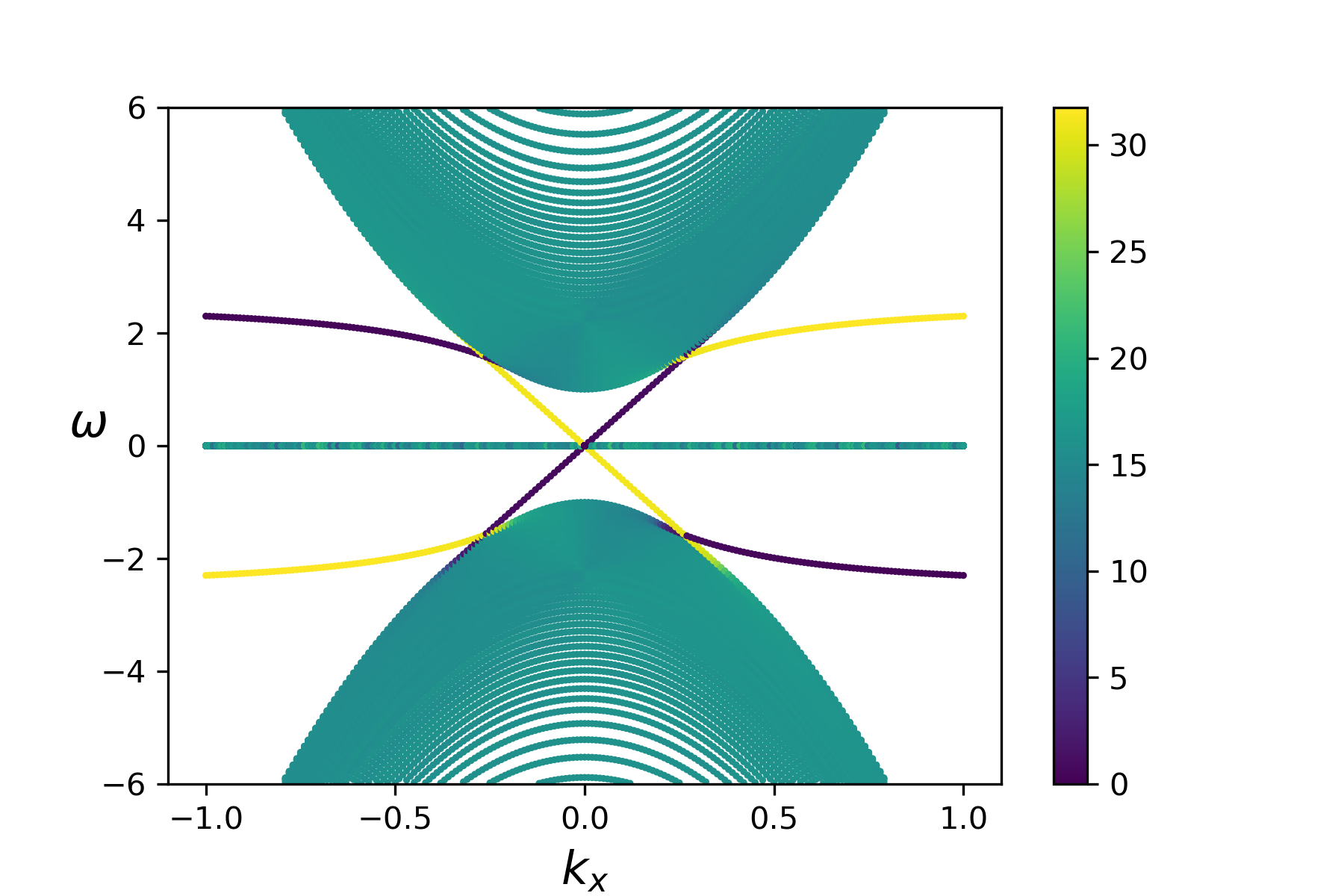}}
	\subfloat[$\gamma=0.156$]{\includegraphics[width=8cm]{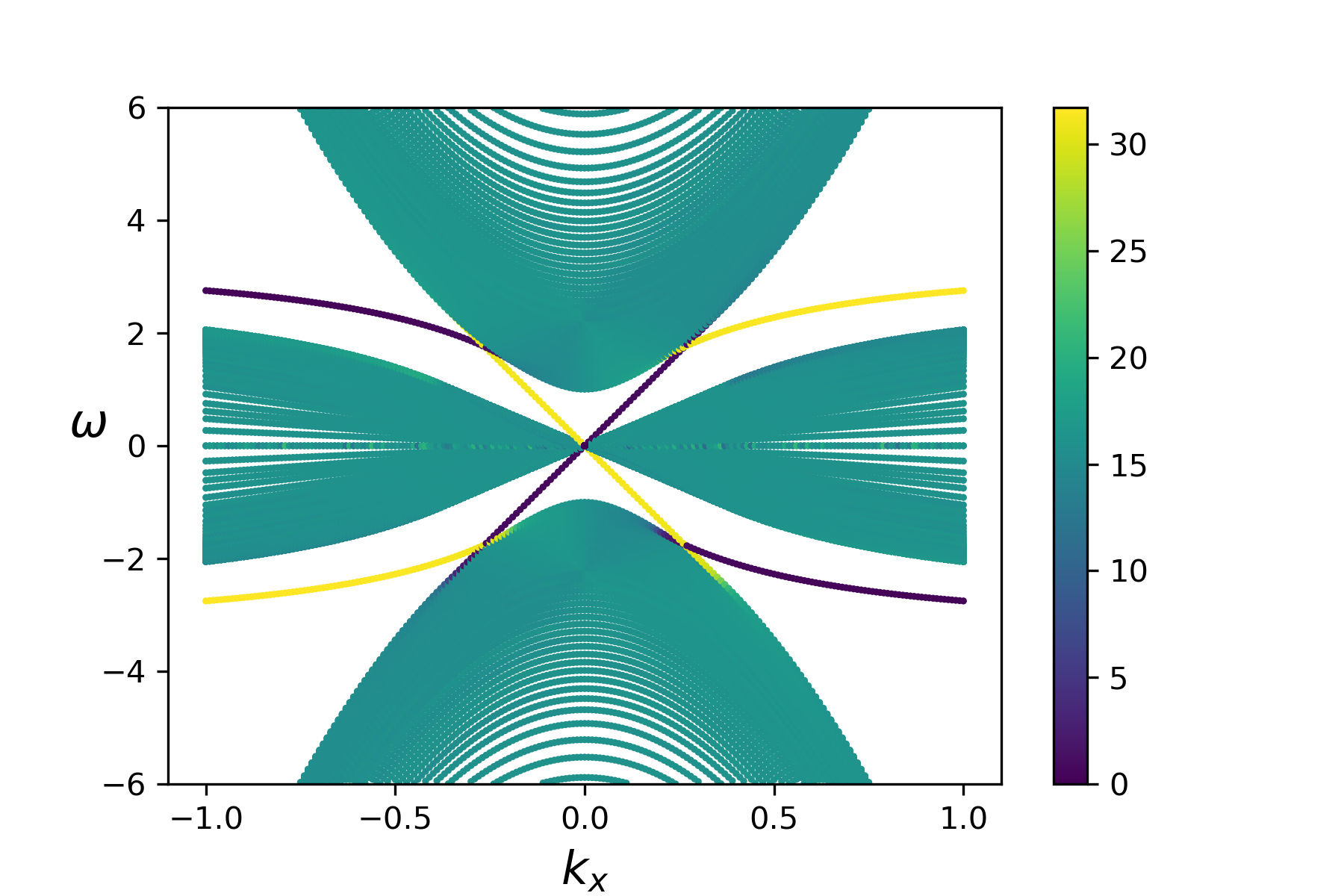}} 
    \caption{Spectrum for (a) shallow water problem with edges with no-slip boundary conditions and (b) for the MHD shallow water problem. We took $\epsilon=0.2$ and strip width $L=32$. The coloring gives the localization of the mode along the strip width. In particular dark blue colour indicates that the modes are localised on the left edge while yellow indicates that the modes are localised on the right edge. Green colour indicates that the modes are bulk modes, not localised on any edge.}
    \label{fig:subfigures}
\end{figure*} 
To further establish the bulk-boundary correspondence and the topological properties of the MHD shallow water wave Hamiltonian \eqref{matrixresca12}, it is also helpful to study Eq.~\eqref{matrixresc09812} for an infinite strip geometry as in \cite{delplacereal}. In this case, we do not consider an equator, i.e. we take $m=1$ everywhere, and instead introduce hard walls at the strip edges along the width of the strip. The hard walls are implemented by imposing no-slip boundary conditions. Using the numerical methods of Dedalus \cite{Burns_2020} for solving spectral problems, we obtain Fig.~\ref{fig:subfigures}. Here, we compare the ordinary shallow water problem, for which this spectrum was obtained in Ref.~\cite{delplacereal} and is depicted on the left hand side of Fig.~\ref{fig:subfigures}, with the magnetohydrodynamic case depicted on the right hand side of Fig.~\ref{fig:subfigures}. We see that with or without the magnetic field, there are two solutions localized at each edge. One is the Kelvin wave, which is expected since the edge problem is similar to a coastal problem where a coastal Kelvin mode appears \cite{thomson_1880}. The second is the Yanai wave, which appears at the edge due to odd viscosity \cite{delplacereal}. This further hilights the robust nature of the magneto-Kelvin and magneto-Yanai waves.

\end{document}